\documentclass[oldversion]{aa}  

\pdfoutput=1
\usepackage{amsmath}
\usepackage{graphicx}
\usepackage{txfonts}
\usepackage{url}
\newcommand{\releasedate}{Feb 8, 2008}
\newcommand{\nspectra}{2470}
\newcommand{\ntargets}{1923}
\newcommand{\nassociations}{520}
\usepackage[authoryear]{natbib}

\usepackage{color}
\newcommand{\new}[1]{{#1}}

\begin{document}

\title{The Hubble Legacy Archive NICMOS Grism Data}

\titlerunning{HLA NICMOS Grism Data}

\author{Wolfram Freudling, 
Martin K\"ummel, 
Jonas Haase, 
Richard Hook, 
Harald Kuntschner, 
Marco Lombardi,
Alberto Micol, 
Felix Stoehr, 
\and
Jeremy Walsh}
\authorrunning{Freudling et al.}

\institute{Space Telescope -- European  Coordinating Facility}
\date{submitted May 29, 2008; accepted Sep 3, 2008}

\abstract{ {The Hubble Legacy Archive (HLA) aims to create calibrated science
data from the Hubble Space Telescope archive and make them accessible via
user-friendly and Virtual Observatory (VO) compatible interfaces.  It is a
collaboration between the Space Telescope Science Institute (STScI), the
Canadian Astronomy Data Centre (CADC) and the Space Telescope - European
Coordinating Facility (ST-ECF). Data produced by the Hubble Space Telescope
(HST) instruments with slitless spectroscopy modes are among the most difficult
to extract and exploit. } {As part of the HLA project, the ST-ECF aims to
provide calibrated spectra for objects observed with these HST slitless modes.
} {In this paper, we present the HLA NICMOS G141 grism spectra.  We describe in
detail the calibration, data reduction and spectrum extraction methods used to
produce the extracted spectra.  The quality of the extracted spectra and
associated direct images is  demonstrated through comparison with near-IR
imaging catalogues and existing near-IR spectroscopy.} { The output data
products and their associated metadata are publicly available
(\url{http://hla.stecf.org/}) through a web form, as well as  a VO-compatible
interface that enables flexible querying of the archive of the \nspectra\
NICMOS G141 spectra. \new{In total, spectra of \ntargets\  unique targets
are included.}}  }

\maketitle

\section{Introduction}

Three of the current Hubble Space Telescope instruments include built-in
slitless spectroscopic modes: the Space Telescope Imaging Spectrograph
(STIS), the Near Infrared Camera and Multi-Object Spectrometer (NICMOS), and the
Advanced Camera for Surveys (ACS).  The main advantage of slitless spectroscopy
is that spectra can in principle be obtained from all objects within the field
of view of an instrument. The main disadvantages are that the size of the
object limits the achievable spectral resolution, that spectra might overlap,
and that the background is relatively high because no slit mask prevents the
full sky background from illuminating every pixel of the detector. In addition,
extracting one-dimensional spectra from such data is a complex process and is
often achieved using highly interactive procedures.  

A fraction of the slitless data in the HST archive has been collected to obtain
spectra of specific objects, but spectra of objects other than the primary
targets have in most cases not been extracted or analysed.  The goal of the
ST-ECF HLA project is to extract spectra of all objects that have been
observed with HST slitless spectroscopy modes and to serve these spectra
through an archive with associated descriptions of the spectra, such as how
much one spectrum is contaminated by those of nearby objects. 

Each of the HST slitless spectrograph modes provides a different capability to
HST.  STIS and the ACS prisms cover the UV part of the spectrum that is not
accessible from the ground. The NICMOS grisms cover IR wavelengths that can
partially be observed from the ground, but where the background is much lower
in space. The ACS optical channels covering the optical wavelength range also
benefit from a combination of lower background and higher spatial resolution
from space. \new{The STIS grating provides the highest  resolving power among
the HST instruments with slitless spectroscopy modes. }

The data analysis for each of the spectrographs presents a significant
challenge, and specialised, mostly interactive tools are available for
individual instruments such as NICMOS \citep{nicmoslook}. For ACS a set of
non-interactive software tools, called aXe, has been developed which, on the
basis of a catalogue of the objects on the associated direct image, extracts
one and two-dimensional spectra \citep{axe1}. This software package was, for
example, extensively used for extracting ACS Wide Field Camera slitless spectra
in the Hubble Ultra Deep Field (HUDF) \citep{axe2}.  For the HLA project, we
have produced a pipeline that is designed to extract spectra from large numbers
of datasets in an unsupervised manner and can be tailored to particular
instruments.

We chose to start the slitless spectroscopy HLA project with the NICMOS
G141 grism dataset. After a brief description of the available data in the
archive (Sect.~\ref{sec:data}), the basic reduction steps applied to the NICMOS
images are described in Sect.~\ref{sec:images}, followed by a detailed
discussion of the spectrum extraction procedure (Sec~\ref{sec:extract}). In
Sec~\ref{sec:cal}, we show a new set of NICMOS spectrum calibration data.

Our extraction pipeline, PHLAG, is described in (Sect.~\ref{sec:phlag}).  The
extensive set of metadata that allows the archive of spectra to be served
through Virtual Observatory (VO) interfaces is summarised in
Sect.~\ref{sec:metadata}.  The calibrated spectra were subject to quality
assessment in terms of their astrometric and spectrophotometric properties and
shown by internal and external comparisons to be well-calibrated (see
Sect.~\ref{sec:qc}).  

\begin{figure}[t]
\includegraphics[scale=0.48]{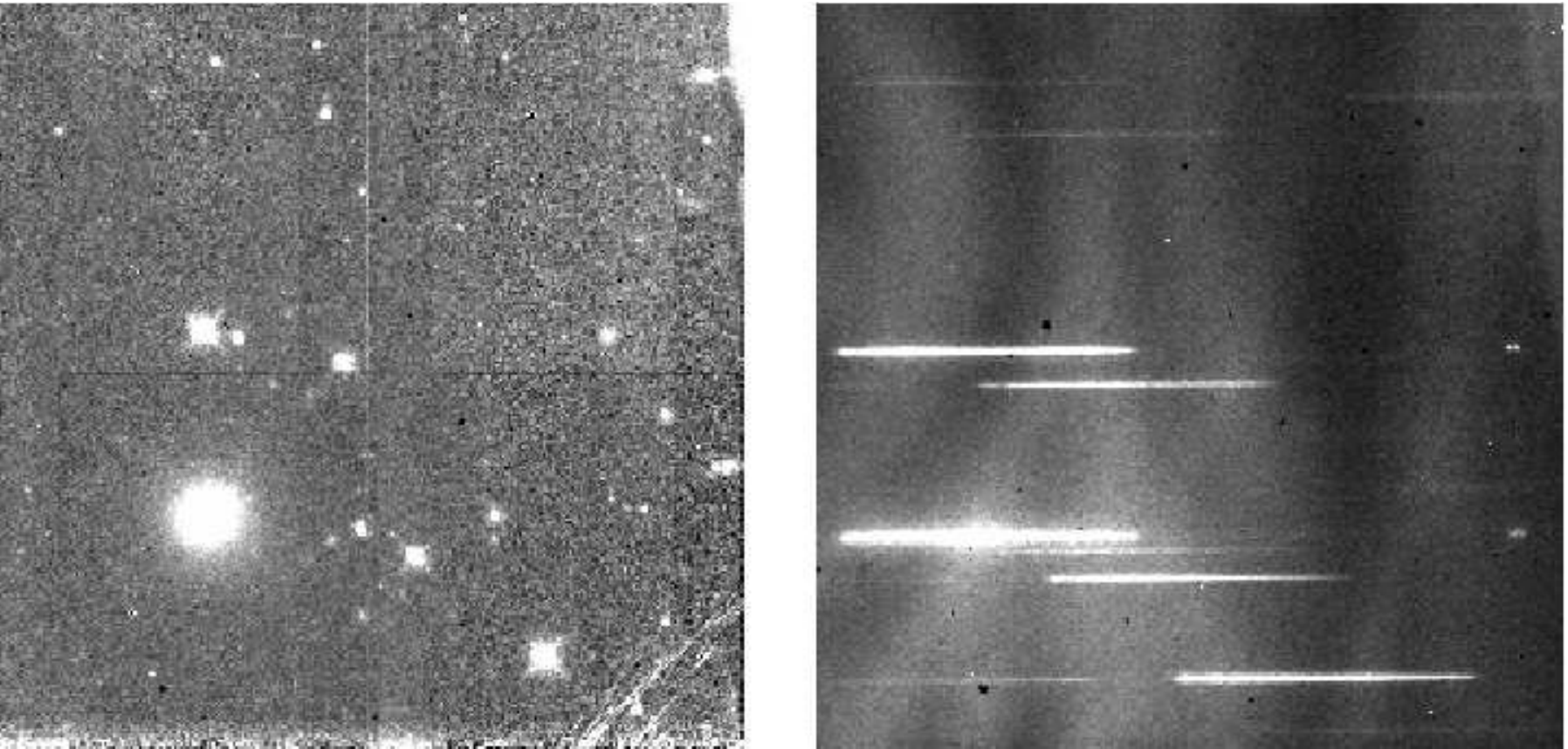}
\caption{Example of one matching pair of a F160W image (left, image n4k6j1a4q)
and the corresponding G141 grism image (right, image n4k6j1zyq)  from HLA
dataset N4K6IZZCQ }\label{fig:grismexample}
\end{figure}

\section{NICMOS Grism Data}\label{sec:data}

NICMOS, a second generation HST instrument, was installed in 1997. One of the
three NICMOS cameras, NIC3 includes three grisms for slitless spectroscopy. A
grism combines a grating and a prism in such a manner that it  produces a
dispersed spectrum at the position of the undispersed image when the grism is
removed. At the nominal grism wavelength, the image through a grism will be, to
first order, identical to an image without a dispersive element.

In NICMOS, the grisms are mounted on a filter wheel that can switch between any
of the filters and grisms. Since there are no slit masks, spectra of all
objects in the fields are produced. Depending on the location of the object,
the zeroth, first or the second order spectra may be visible. The mode of
operation is to first take an image with a filter that matches the bandpass of
the grism followed by one or more exposures with the grism. To avoid image
defects, a small or large scale dither pattern is used in a well designed
observing run. Fig.~\ref{fig:grismexample} shows a typical pair of exposures of
a filter image with a matched grism image.

The most widely used of the NICMOS grisms is G141, which covers the wavelength
range from 1.10 to 1.95$\,\mu$m. This wavelength range is not easily accessible
from the ground, and the dark background in space makes G141 a sensitive mode
to obtain spectra of all objects within the 51 $\times$ 51 arcsec$^2$ field of
view of the camera. The dispersion  for the first order spectrum is about 8.0
nm per pixel, which results in spectra about 105 pixels long. 

In March 2002, the NICMOS Cryocooler System (NCS) was installed. With the NCS,
the NICMOS detectors operate at a higher temperature and this changed the dark
current as well as the quantum efficiency as a function of wavelength for each
pixel. Therefore, all calibrations have to be derived separately for data taken
with and without the NCS \citep{ncscal}.

\begin{figure}[t]
\includegraphics[scale=0.4]{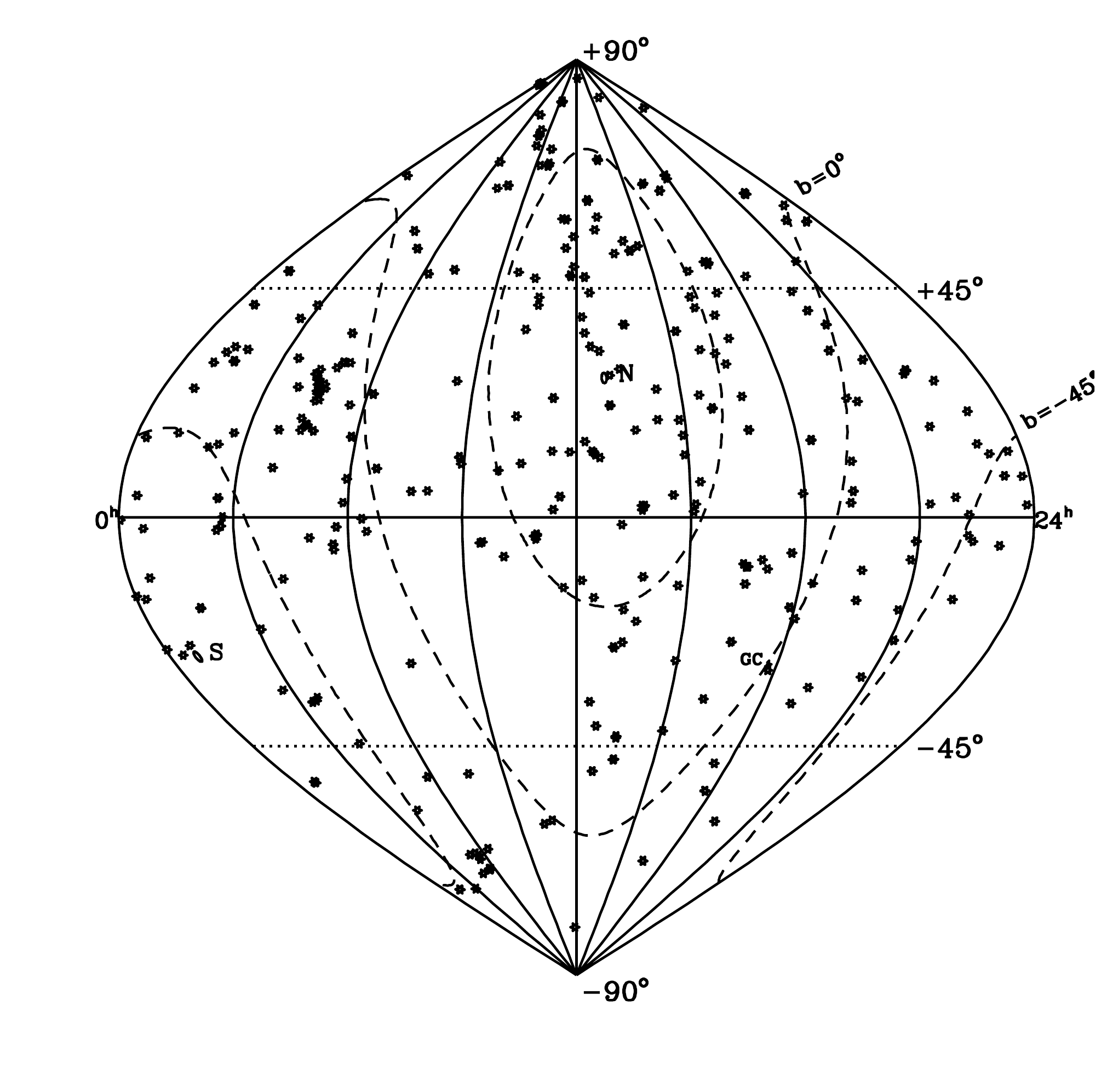}
\caption{Distribution of NICMOS G141 grism images on the sky.}\label{fig:sky}
\end{figure}

In July 2007, the HST archive contained  about 9000 NICMOS G141 observations.
\new{These data include both pointed observations towards specific targets and
programmes which used NICMOS in parallel to other HST instruments.  More than
50\% of the NICMOS data presented in this paper are from four different HST
programmes,  namely HST proposal IDs 9707 \citep{john1}, 9865 \citep{mal1},
8082 \citep{john2} and 10226 \citep{mal2}.} The pointing positions are almost
randomly distributed on the sky and are shown in Fig.~\ref{fig:sky}.

\section{Image Processing and Target Selection}\label{sec:images}

\subsection{Associations}

Each spectrum in this work is based on a collection of images, which includes
both grism images and undispersed images.  The images in each of these {\em
associations} overlap and were treated collectively. The undispersed images
were co-added for the creation of target lists, and the two-dimensional spectra
on the grism images were rectified and co-added before the extraction of
one-dimensional spectra. These steps are described in detail in this section.

The grouping of NICMOS datasets into associations was performed using the World
Coordinate System (WCS) in the image header to compute relative pointings. To
ensure accurate relative coordinates among the images in any given association,
only datasets obtained with the same set of guide stars were included.

The spectral dispersion of NICMOS grism images is along the x-axis of the
detector.  The orientation of the grism images on the sky determines which of
the spectra overlap. We therefore chose to co-add only grism images that differ
in orientation by less than one degree, whereas the orientation of undispersed
images was not restricted. \new{The result of this procedure is that, for some
targets, several spectra are extracted and not co-added. The reason for this
decision was that different orientations of the spectra result in different
contamination. Indeed, the very reason that many programmes have observed the
same field with different roll angles is that it is likely that at least one of
the spectra for each target is free of contamination. We leave it up to the
user to pick the best of the spectra for each target.   }

In total 962 NICMOS G141 associations were created from 9262 grism members (up
to June 2007).  The grouping efficiency is therefore more than 9.5 exposures
per association.

\subsection{Preprossessing of NICMOS Images}

To produce one-dimensional spectra of individual targets, we extracted the flux
and wavelength information from NICMOS grism images and associated undispersed
images that have been reduced with STScI's CALNICA reduction pipeline version
4.1.1 \citep{nicmoshandbook}.  NICMOS supports non-destructive multiple
readouts, and virtually all NICMOS data are taken in this mode.  The STSDAS
program CALNICA \citep{calnica} produces calibrated images from these raw data,
the processing steps include bias correction, dark current subtraction, and
computation of the count rate for each pixel using all readouts. The NICMOS
images suffered from a number of peculiar anomalies, some of them are treated
by CALNICA, for others separate STSDAS tasks are available.  The default
parameters of CALNICA and some of the algorithms used in the auxilliary task
are optimised for use with undispersed images. In the rest of the section, we
describe the parameters and procedures we used for the NICMOS Grism HLA.

\subsubsection{Bars Correction}

Some NICMOS images suffer from pairs of bright and dark columns or rows,
so-called ``bars''.  \new{These bars are bias-related artefacts in the form of
noiseless offsets} of a few DNs along a pair of columns or rows, with the
pattern replicated exactly in all four image quadrants. CALNICA has a built-in
procedure to detect and remove such bars.  Unfortunately, the central two rows
of NICMOS spectra contain more than 80\% of the flux of point sources and the
spectra are therefore of similar width to the bars. As a result, in crowded
grism images when the spectra are well aligned with the rows of the detector,
CALNICA often mistakes the peak of the spectra for bars and tries to remove
them.  To avoid this problem, a value of 4$\sigma$ was used for the threshold
in CALNICA which controls the bar detection. This compares to a default value
of 3$\sigma$.  It was found that this selection removes the described problem
in virtually all cases.

\subsubsection{Bad Pixels}

Images reduced with CALNICA typically contain between 20 and 100 unflagged bad
pixels, which contain significantly higher flux than their neighbours. These
``hot'' pixels can produce spurious emission lines in extracted spectra.  The
pixel spacing of the NICMOS NIC3 camera is 0.2$\,$arcsec which undersamples the
Point Spread Function (PSF).  Commonly used methods to distinguish between
point sources and hot or cold pixels, based on the sharpness of features on the
calibrated image, are therefore not very effective with NICMOS.  On the other
hand, the multiple readouts provide information that can be used to recognise
unreliable pixels. The following procedure was implemented to identify bad
pixels. For each pixel the accumulated count rate at each readout, which is one
of the outputs of CALNICA, was fitted by a line. A pixel was flagged as ``bad''
when either the slope of the line was different from zero by at least four
times the uncertainty of the slope, or the reduced $\chi^2$ of the fit was
greater than four.  Flagged pixels were ignored in the subsequent analysis.

\subsubsection{Cosmic Rays}

Because of the multiple readouts, the count rate of a pixel can be measured
even if that pixel is hit by a cosmic ray (CR). The extra charge deposited by
the CR between two readouts produces a sudden discontinuity in the count rate.
CALNICA includes a procedure to identify and remove such jumps from the data.
This procedure is effective for more than 90\% of all CR hits. \new{ CALNICA
version 4.1.1. sometimes fails and produces a bright pixel in the output
images, even though it correctly recognised that this pixel is affected by a
CR. A future release of CALNICA will fix this problem \citep{newcalnica} but is
currently not yet available. } Because of the undersampling of the NICMOS
images, distinguishing between stars and cosmic rays cannot be easily done on
the basis of the sharpness of a flux peak alone. We therefore adopted a
procedure to use both the charge built up and the sharpness to minimize the
number of false detections of CR hits.  Each pixel identified by CALNICA as
affected by CR was checked to determine whether this pixel had an unusually
high flux in the output images.  Specifically, such a pixel was flagged as bad
if its flux was more than 4$\sigma$ above the mean of its neighbouring pixels.

\subsubsection{Pedestal}

NICMOS images suffer from random DC offsets in the bias level, which may be
different in each of the four quadrants. The offsets are constant within each
quadrant. If not removed, these offsets imprint the structure of the flatfields
on the calibrated images during the flatfielding step. Since grism images are
not flatfielded as images, the pedestal is visible as different constant
offsets of the  quadrants. There are two STSDAS tasks, pedsky and pedsub,
available to remove the pedestal from the calibrated images.  Neither of them
works well with grism images. We therefore used the following approach to
improve the pedestal in the grism images. First, for each pixel of a quadrant
we normalised the count rates of the multiple readouts to the mean of the last
three reads. We then computed the median count rate for each readout of all the
pixels in a quadrant. We then computed the value of the DC offsets for each
quadrant that results in constant median count rate. These values were then
subtracted from all the pixels in the corresponding quadrant.

\subsection{Co-addition of Undispersed Images}\label{sec:coadd}

In order to create a deeper and cleaner direct image, to be used to prepare the
input object catalogue needed as input to the spectral extraction, the direct
images in each association were first combined.  The MultiDrizzle software
\citep{drizzle,Fruchter} was used to register each input image, based on its
header WCS, and to remove the small NICMOS geometric distortion using the
standard cubic polynomial coefficient set. Standard default settings were used.
The data quality arrays were also aligned by this step and registered output
weight images created. MultiDrizzle is not yet well optimised for NICMOS and
experiments revealed that a combination of inadequately precise alignment and
the strong undersampling of the NIC3 camera resulted in the cores of compact
objects being incorrectly flagged as cosmic rays during the ``driz\_cr''
MultiDrizzle step. This led to significant photometric errors. To avoid this
\new{ we used MultiDrizzle only up to the image combination step. } The
preparation of image bad pixel masks and the final image combination were
performed using the {\it imcombine} task in IRAF.

\subsection{Target Catalogue}
\label{sec:undisp-image-catal}

The first step in the extraction of slitless spectra is to find objects  on the
undispersed image.  Object parameters relevant for the extraction are the
coordinates, the position angle, the object sizes, and the magnitude in a
reference spectral band.

\begin{table*}
  \caption{Relevant parameters used for the source extraction in the direct images.}
  \centering
  \begin{tabular}{lcl}
  \hline
    Parameter & Value & Description \\
    \hline
    \texttt{DETECT\_MINAREA} & $4$ & Minimum number of pixels above
    threshold \\
    \texttt{DETECT\_THRESH} & $3.0$ & Minimum threshold for detection
    \\
    \texttt{ANALYSIS\_THRESH} & $3.0$ & Minimum threshold for analysis
    \\
    \texttt{DEBLEND\_NTHRESH} & $8$ & Number of deblending sub-thresholds
    \\
    \texttt{DEBLEND\_MINCONT} & $0.03$ & Minimum contrast parameter for
    deblending \\
    \texttt{SEEING\_FWHM} & $0.26$ & Stellar FWHM in arcsec \\
    \texttt{BACK\_SIZE} & $64$ & Background mesh size \\
    \texttt{BACK\_FILTERSIZE} & $1$ & Background filter size \\
  \hline
  \end{tabular}
  \label{tab:sex}
\end{table*}

We used the SExtractor program \citep{ba} to generate the object catalogues.
The output exposure time image generated in the co-addition step was used as a
weight map, which significantly  reduces the number of spurious detections and
improves the accuracy of the photometry.

The relevant parameters used for the extraction are given in
Tab.~\ref{tab:sex}.  We used the ``windowed'' centroid (\texttt{XWIN},
\texttt{YWIN}) as the target coordinates, which is the  iteratively computed
first moment of  the object's surface brightness after convolution with a
Gaussian function with size matched to the object.  This leads to very accurate
centroid measurement, especially for point sources (the large majority of
sources in our sample).  In addition, we also measured ``un-windowed''
morphological major axes \texttt{A}, \texttt{B}, position angle \texttt{THETA},
and  magnitudes (\texttt{MAG\_AUTO}.  Finally, we kept track of the success of
these measurements by retaining the SExtractor \texttt{FLAGS} parameters (see
below).

The original SExtractor catalogue was then further filtered in order to remove
spurious detections:

\begin{itemize}

\item We removed all objects with $\mathtt{FLAGS} \ge 64$: that effectively
removes all objects for which a severe memory overflow occurred either during
deblending or during the extraction;

\item We removed faint objects for which we do not expect to be able to extract
a spectrum from the dispersed images ($\mathtt{MAG\_AUTO} \ge 23.5$), or
extremely elongated ones ($\mathtt{A} / \mathtt{B} > 4$), which are often
simply detector artifacts or cosmic rays;

\item We took objects with sizes smaller than one pixel ($\sqrt{\mathtt{A}
\mathrm{B}} < 1 \mbox{ pixel}$) as point-like sources, and set both their axes
to $1 \mbox{ pixel}$;

\item For extended objects, we computed the extension and orientation of the
virtual slits from the parameters \texttt{A}, \texttt{B}, and \texttt{THETA}
(see Appendix~A);

\item We discarded associations that suffered from significant crowding ($>140
\mbox{ sources arcmin}^{-2}$) \new{since such a high source density leaves too
little  sky background to allow the estimation of the local background levels.}

\item We included external 2MASS sources in the final catalogue, as described
in Sect.~\ref{sec:extern-2mass-sourc}.

\end{itemize}

\subsection{Astrometric calibration}
\label{sec:astr-calibr}

The astrometric coordinate system  of the raw NICMOS images is specified in the
WCS keyswords in the image headers.  Its accuracy  is ultimately  limited by
the accuracy of the catalogue used for pointing the telescope. For earlier
data,   the Guide Star Catalog~I \citep{1990AJ.....99.2019L} was used, and
since 2000 the Guide Star Catalog II \citep{gsc2} is used.  Both catalogues are
based on scans of the photographic Sky Survey plates at various epochs and
bandpasses.  The absolute accuracy reaches $\sim 0.3''$ over a large fraction
of the sky, but  errors can be as high as several arcsecs towards the edges of
the scanned plates.  A comparison of the astrometry in undispersed NICMOS
images with more accurate data showed such large offsets in some cases.

Accurate coordinates are not only important for the final astrometric fidelity
of our spectra, but also to identify potentially contaminating objects outside
of the NICMOS images (see Sect.~\ref{sec:contam}). We therefore decided to
carry out an independent astrometric calibration of all our undispersed images.
For that purpose, we first generated  preliminary versions of the  catalogues
described in Sect.~\ref{sec:undisp-image-catal} with coordinates based on the
WCS and used them to compute offsets to the WCS.

\begin{table*}
\caption[]{Reference Catalogues used for Astrometric Calibration.}\label{tab:refcats}
\centering
\begin{tabular}{lccc}
\hline
Catalogue & Acronym & Reference & Accuracy \\
\hline
USNO CCD Astrograph Catalog& UCAC2 & \cite{2004AJ....127.3043Z} & 0.06
arcsec\\
Two Micron All Sky Survey& 2MASS  &\cite{2006AJ....131.1163S} & 0.10 arcsec\\
Sloan Digital Sky Survey& SDSS-DR5  &\cite{2007ApJS..172..634A} & 0.15
arcsec \\
USNO-B1.0 catalog & USN0& \cite{2003AJ....125..984M} & 0.20 arcsec\\
Guide Star Catalog II & GSC2.3.2 & \cite{gsc2} & 0.30 arcsec\\
\hline
\end{tabular}
\end{table*} 

The offset correction applied to the WCS is the mean offset of the WCS
catalogue coordinates and a global astrometric reference catalogue obtained by
merging the five catalogues listed in Tab.~\ref{tab:refcats}.  The merged
astrometric catalogue includes sources from all reference catalogues and
assigns to each object  the coordinates of  the most accurate catalogue in
which it was detected.  The mean offset was then calculated using a weighted
average, with weights proportional to the inverse of the square intrinsic
astrometric errors associated with each source.

The most critical part of the whole procedure is the matching of the sources in
the astrometric catalogue with our object catalogue.  This step is challenging
because of the small field of view of the NIC3 camera and the relatively large
number of artefacts present in the NICMOS images.  The basic approach of our
adopted procedure is to first use the WCS coordinates and identify the nearest
neighbours in the merged reference catalogues, compute offsets to the WCS, and
iterate. The scatter in the offsets of different stars is then used as an
estimate of the astrometric accuracy.

\begin{figure}[t]
\includegraphics[width=\hsize]{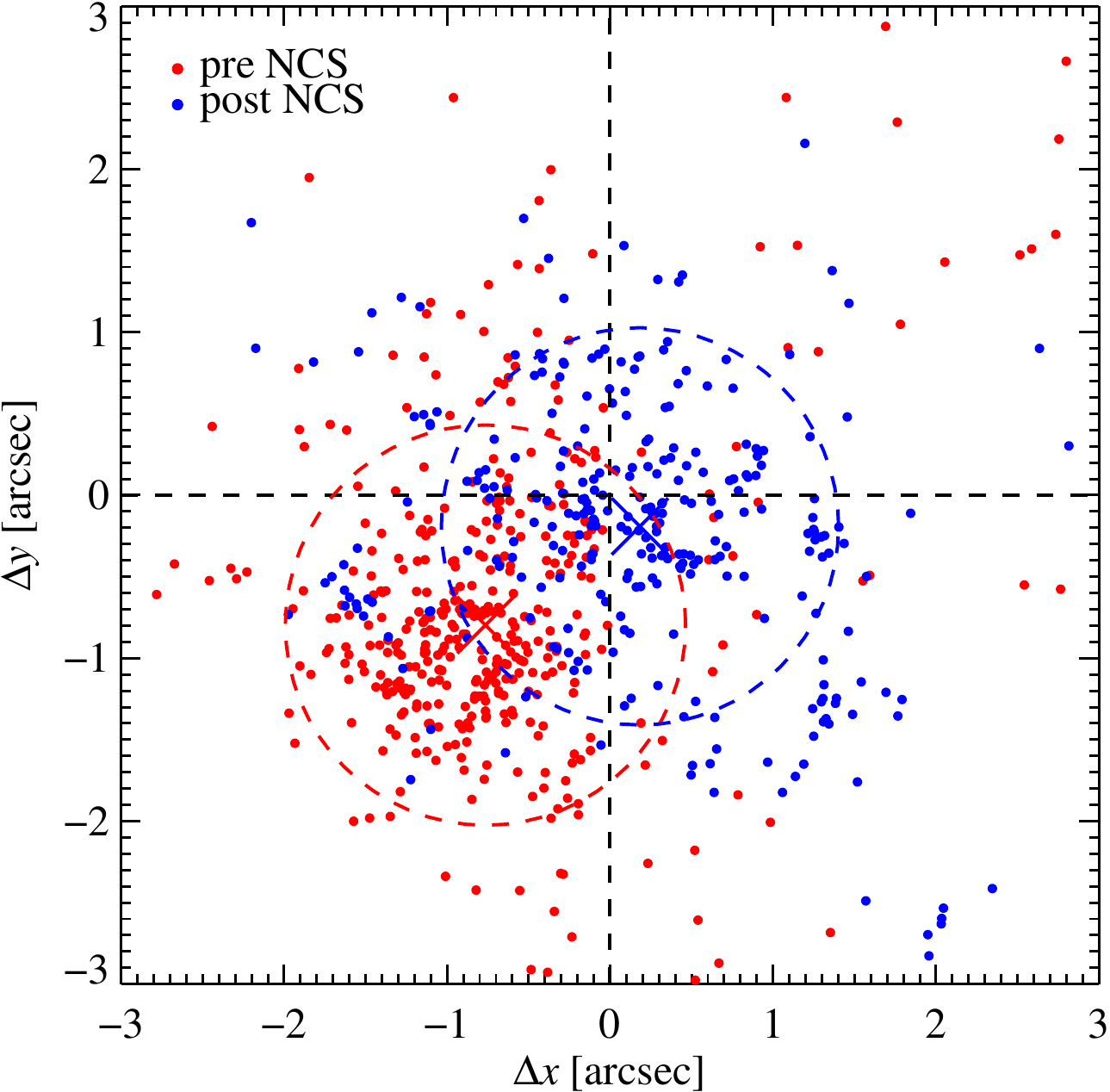}
\caption{Difference between objects coordinates computed from  the WCS in the
image header and coordinates listed in astrometric reference catalogues. The
crosses mark the mean of the distributions before and after the installation of
the NCS, and the circles make the 1$\sigma$ scatter around the mean. The
systematic error in the WCS coordinates is significant larger for the pre-NCS
data.  }\label{fig:astro_offsets} 
\end{figure}

Our method, however, differs in a number of ways from the simple scheme
outlined above.  Firstly, the initial astrometric solution assumed is not the
original HST astrometry, but rather the HST astrometry corrected for the median
offset computed among all the pre- or post-NCS observations. This median offset
is $\sim 1.1''$ for pre-NCS observations, and $\sim 0.3''$ for the post-NCS
observations (see Fig.~\ref{fig:astro_offsets}).

Secondly, we used the \textit{median\/} offset between the astrometric
catalogue and the NICMOS object catalogue in the first two iterations, and a
$\sigma$-clipped weighted average for subsequent iterations.  This ensures that
the initial, highly uncertain offset does not introduce any strong bias because
of wrongly matched pairs.

Thirdly, if during any iteration we found more than three matched pairs, we
uses a clustering analysis to remove spurious matches.  Specifically, we
identified clusters in the offset plane $(\mathrm{d}x, \mathrm{d}y)$.  We then
removed from each cluster those matches that increase the standard deviation of
the offsets by more than a factor of two.  This technique is more robust in
removing badly matched pairs than $\sigma$-clipping.

\begin{figure}[t]
\includegraphics[width=\hsize]{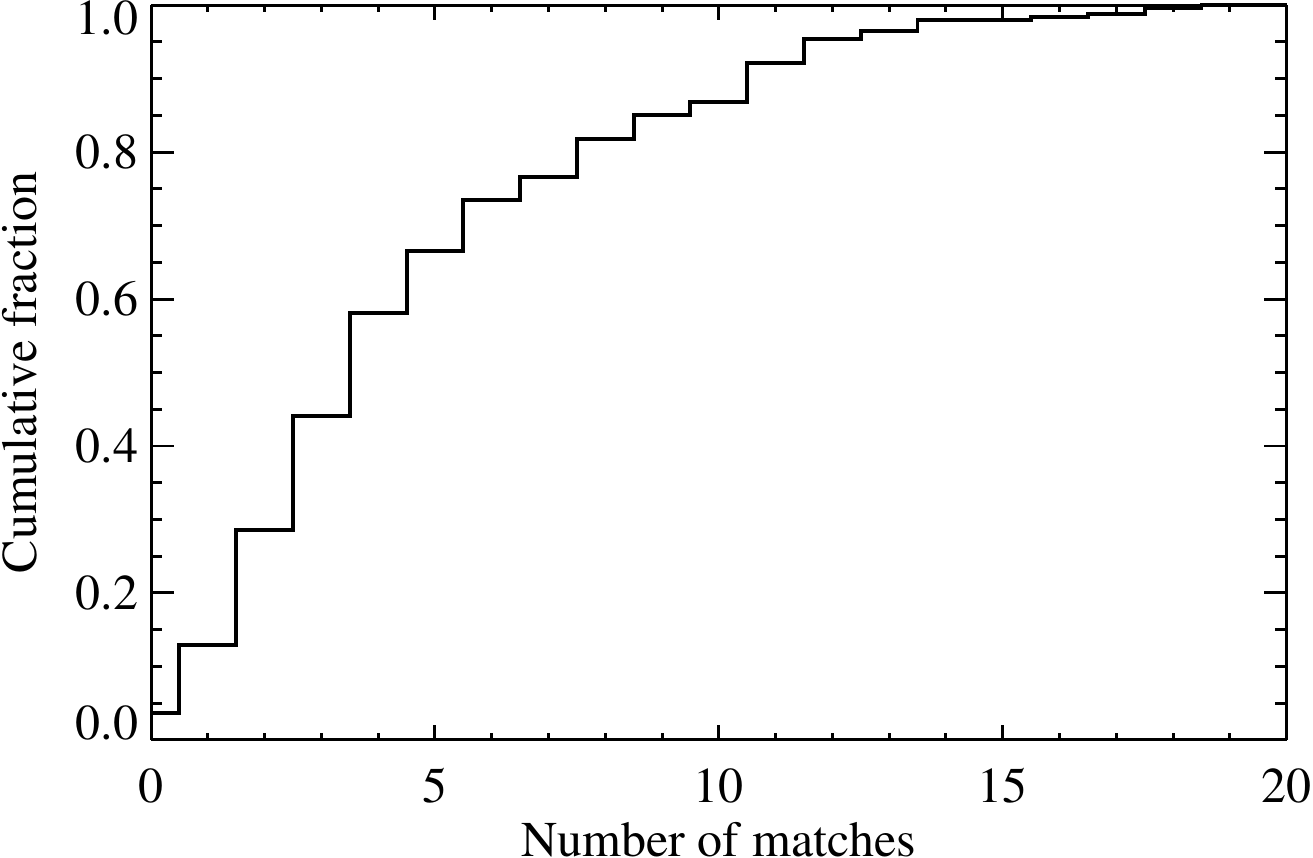}
\caption{Cumulative distribution of number of astrometric calibration stars
used in each field.}\label{fig:21}
\end{figure}

\begin{figure}[t]
\includegraphics[width=\hsize]{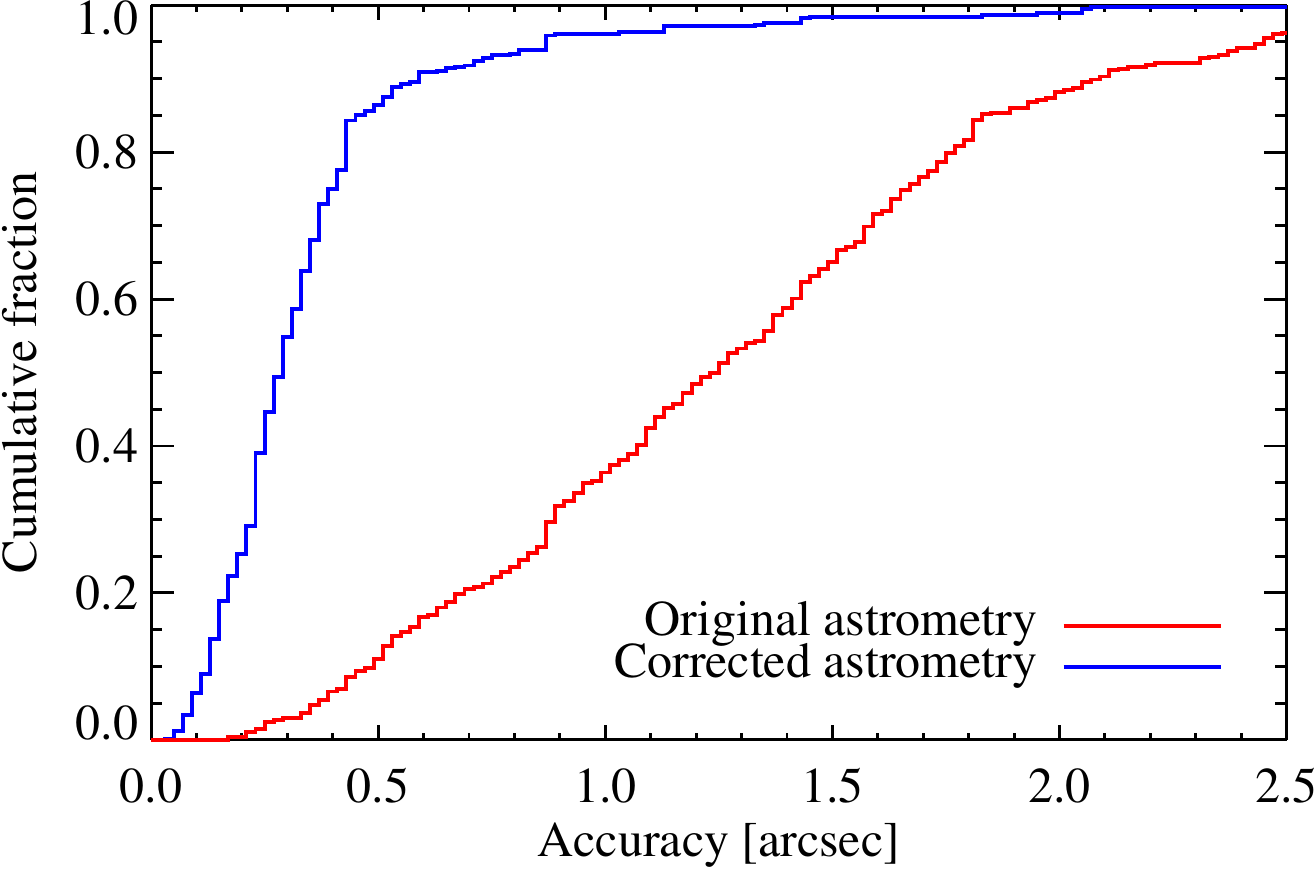}
\caption{ Cumulative distribution of estimated astrometric accuracies of the
object coordinates before and after the astrometric correction.
}\label{fig:22}
\end{figure}

Finally, in the case of the 2MASS catalogue, we also compared H-band magnitudes
measured on the NICMOS images with those in the 2MASS catalogue and removed bad
matches.

At the end of the whole process the algorithm returned the following quantities:

\begin{itemize}

\item the estimated offset;

\item the theoretical statistical error on the offset, derived from a simple
error propagation of the intrinsic catalogue accuracies listed in
Tab.~\ref{tab:refcats};

\item in cases where at least two objects are matched, the sample variance of
the offsets of the matched objects was also calculated;

\item finally, the full list of matches was retained for inclusion in the
metadata.

\end{itemize}

In total, about 96\% of our released spectra have corrected coordinates, and
for 91\% of these we were able to match two or more astrometric sources The
distribution of the number of matches is shown in Fig.~\ref{fig:21} The
computed offsets are shown in Fig.~\ref{fig:astro_offsets}.  In
Fig.~\ref{fig:22} we compare the  distributions of the estimated astrometric
uncertainty (i.e.\ the square root of the sample variance of offsets) before
and after applying these offsets. The median  accuracy of the astrometry
improves from about 1.24 arcsec to about 0.27 arcsec.

\subsection{External 2MASS sources}
\label{sec:extern-2mass-sourc}

NICMOS grism images can include partial spectra of bright objects that are
located outside of the field of view (FOV) but close enough to the edge of the
image so that the grism still deflects part of the first order spectrum onto
the detector.  The zeroth order spectra of such objects can be fully included
in the grism image without a corresponding counterpart on the undispersed
image.  Such partial spectra might contaminate other spectra, and the zeroth
order spectra might be mistaken for bright emission lines.  We used the 2MASS
catalogue to identify bright sources close to each grism image that might
create such spurious spectra.  After the astrometric correction described in
Sect.~\ref{sec:astr-calibr}, the coordinates of the images are known well
enough to predict the location of such spectra accurately enough to take them
into account for contamination estimation.  By visual inspection of  the images
we determined that the maximum distance of sources from the image edges that
still produces a spectrum is 13 arcsec at the left image edge, and 3 arcsec at
the right image edge.

Although it is in principle possible to extract spectra of the external
sources, in the current release we did not attempt to do so, but only used the
external sources for the estimate of the contamination see
Sect.~\ref{sec:contam}.

\section{Extraction of NICMOS Spectra}\label{sec:extract}

\subsection{Methods}

One-dimensional integrated \new{first order} spectra of objects were extracted
from grism data with the help of the target catalogues. \new{Whilst second
order spectra are visible on NICMOS grism images for bright targets, in most
cases only small parts of these spectra are usable because they are not
completely contained within the NICMOS detector and usually suffer from
significant contamination.} We therefore chose to limit our project to the
first order spectra.

\new{The essential extraction steps} were to remove the local background close
to the spectrum, use the position of the object on the undispersed image to
obtain a wavelength scale, and then add up the flux values for each wavelength
bin. The wavelength dependent response of  each pixel was taken into account
during this extraction.

We used a modified version of the extraction software aXe \citep{axe1} version
3.7 for that purpose.  aXe was designed to treat slitless spectroscopic data
and includes the transfer of source positions derived on undispersed images
onto the slitless images; the application of wavelength dependent  flat fields,
the co-addition and extraction of spectra and the  estimation of the
contamination. We now describe the modifications and additions that were made
to the basic aXe package, most of them address NICMOS specific properties of
the data. 

\subsection{Tracing of spectra} \label{sec:spectraces}

Spectra on G141 grism images are closely aligned with the rows of the detector.
The angle between the dispersion direction  and the detector rows depends on
the exact orientation of the grisms.  The grism is mounted on a rotating filter
wheel, which lead to slight differences in the positioning of the grisms each
time it is rotated into place. Typical values for the angle between the
dispersion direction and the x-axis of the detector ranged from 0 to 2$^\circ$.
We therefore determined this angle separately for each grism observation.

To ensure an identical extraction around the variable traces for the entire
data set, the spectral trace was measured on each grism image individually.
This was done by selecting the brightest three point-like objects on each image
and determining the trace solution for each object individually. The individual
solutions were then combined using the signal-to-noise ratio of the spectrum as
weight to determine the single trace solution for that grism image.

\subsection{Background Subtraction}

The NICMOS grism images are  not flatfielded in the standard pipeline, because
the local quantum efficiency is wavelength dependent. Spatial variations in the
throughput were therefore taken into account during the extraction process for
each target spectrum, after the wavelength calibration. An imprint of the
flatfield convolved with the spectrum of the sky background in the grism
passband is therefore visible in calibrated grism images. This background has
to be {\em subtracted} before the flux in each wavelength bin is summed. In
addition to the imprint of the flatfield, a region of enhanced dark current was
visible in the lower right-hand part of all NICMOS images. This enhanced dark
current, which varies depending on the positioning of the field offset mirror
(FOM), is also an additive component that has to be subtracted.

G141 background images were constructed from grism images that contained no or
few visible spectra, using about twenty different images for each prepared
background.  The median of the scaled images was subtracted from each
individual image to isolate the contribution of the FOM to the background.  In
this manner, two different background images were prepared for different time
periods, one presenting the flatfield imprint on the background, and the other
the enhanced dark current.  Scaled versions of these backgrounds were
subtracted from each grism image.

The scaling of the subtracted background was a crucial parameter for the
flatness of the resulting image and therefore the quality of the extracted
spectra. Because of the uncertainties in the overall bias level and dark
current and the changing sky brightness, the measured background level was not
a good predictor of the structure seen in the background. The scaling for the
subtraction was therefore determined by finding the least-squares solution for
a line fit to  the pixel values in the background versus the ones in the grism
images. Only pixels that do not contain any spectrum were used in this
procedure.

For most data sets, this procedure succeeded in removing most variations in the
background. To further improve the quality of the extraction, a local
background was subtracted  for each extracted spectrum. This local background
is a linear fit to the region around each spectrum.

\subsection{Virtual Slits}

The slit in a longslit spectrograph selects a part of the sky, and light from
that area is then dispersed and binned in wavelength.  In slitless
spectrographs, the light that reaches the final wavelength bins is only limited
by the extent of the targets.  The light distribution of target objects
therefore  defines the lines of constant wavelength on the dispersed image, and
the effective resolving power. The size, position and orientation of the lines
of constant wavelength are used to extract one-dimensional spectra from the
two-dimensional image. 

Target  objects can be of  an arbitrary shape that may depend on wavelength.
For the extraction of spectra, we made two approximations.  Firstly, we
neglected any wavelength dependence of the object shape and derived all shape
parameters from the undispersed image. Secondly, we treated all objects as
ellipses, i.e.  assumed that the isophotes were elliptical.  The spectral
resolution of the spectra extracted with these approximation was slightly lower
than an optimal extraction that uses knowledge of the exact shape of objects as
a function of wavelength. For these approximations, we computed the extraction
direction that optimizes the spectral resolution of the spectra. The details
are given in Appendix~A.

\subsection{Adjustment of Wavelength Scale}

As a result of the filter wheel non-repeatability, the spectra on any given
exposure can be shifted by up to a few pixels relative to the expected
wavelength zero point derived from the undispersed image.  The wavelength
zeropoint was therefore re-adjusted for each exposure by extracting the
brightest few ($\sim$4) spectra and cross-correlating the characteristic
spectral feature (i.e. the drop of sensitivity) at the red end of all spectra
with a fiducial template. This template was prepared from the sensitivity
curves shown in Fig.~\ref{fig:throughput}. In this way any wavelength
zero-point shifts introduced by the non-repeatability of the positioning of the
grism wheel were taken out with a precision of about $\pm$0.2 pixel or $\pm$16
\AA. Potential wavelength calibration changes as a function of position within
the field of view were not taken into account.

\begin{table*}[t]
\begin{center}
\caption{Tabulated values for the linearity correction $b$}
\label{tab:nonlin}
\begin{tabular}{ccccccccccccc}
\hline

$\lambda [\mu]$ m&.825&.875&.925&.975& 1.1&1.2&1.3&1.4&1.5&1.6&1.7&1.8\\
b &.069&.057&.052&.050&.049&.048&.041&.023&.013&.008&.004& 0\\
\hline
\end{tabular}
\end{center}
\end{table*}

\subsection{Pixel Response Function}\label{sec:prf}

The NICMOS detectors show significant sensitivity variations across individual
pixels \citep{sto99,xu03}. These variations, together with the large pixel size
and the small trace angles (see Sec~\ref{sec:spectraces}), result in a
modulation of the extracted spectrum with a wave-like pattern for point-like
objects. The amplitude of this pattern is about 4\%. To correct for this effect
the exact location of the trace must be known to a fraction of the pixel size. 

The subpixel accuracy for the trace location  within a spectrum cannot be
achieved with the fits described in Sec~\ref{sec:spectraces}. A sensitive
estimator of the trace location is the quantity $f = v_{1}/ v_{2}$ with  the
brightest pixel value $v_{1}$ and the second brightest pixel value $v_{2}$,
respectively. For each point-like object, the values $f$ along the spectrum
were determined. The position where this function peaks marks the wavelength
value at which the spectrum is located at the centre of a row.  For raw spectra
with sufficient signal-to-noise to determine the peaks were multiplied by a
correction function~$f_{\rm pr}$:

\begin{equation}
f_{\rm pr} = 1 + 0.04  \sin (2 \pi  y) ,
\end{equation}
where $y$ is the position of the spectrum on the detector determined with
subpixel accuracy as described above.

\subsection{Non-Linearity Correction}

The NICMOS detector responds to different infalling photon count rates in a
non-linear manner. Even if the total number of photons is identical, the
detector will report a different flux if the number of photons per unit time
differs.  This effect was described  in \cite{bohlinI}. Following
\cite{bohlinII}, we corrected point-like objects for the detected photon count
rates $c$ according to

\begin{equation}
c = c_{\rm obs} / [ 1 - 2 b(\lambda)  + b(\lambda)  \log_{10}(c_{\rm obs})], 
\end{equation}
where $c_{\mathrm obs}$ is the count rate derived from the grism image, and
the values $b(\lambda)$ are given in Table \ref{tab:nonlin}.

\subsection{Flux Calibration} \label{sec:flux}

Flux calibration of point sources was done by multiplying the count rate of
extracted spectra by the sensitivity curve (see Fig.~\ref{fig:throughput})
derived from flux standards.  Flux standards are stars and thus point-like
sources, and the spectral resolution is determined by the instrument setup. For
extended objects, however, the spectral resolution is degraded by the extent of
the object in the dispersion direction.  This smoothing of the spectra reduces
the amplitude of structure in the sensitivity curve and has to be taken into
account for the flux calibration of extended objects.

The flux calibration of extended sources was carried out by multiplying the
count rate by a smoothed version of the point source sensitivity function.
The width of the Gaussian smoothing kernel was
\begin{equation}
\sigma = f r\sqrt{s^{2} - p^{2}}  , 
\end{equation}
where $s$ is the FWHM of the object in the dispersion direction, $p$ is the
FWHM of the point spread function, and $r$ is the dispersion in $\AA$ per
pixel. The quantity $f$ is a correction factor that was found empirically by
experiment using extended source spectra; the adopted value was 0.65.

The procedure was tested both on  simulations of Gaussian objects using the aXe
simulation package aXeSIM \citep{axesim}, and on real extracted NICMOS spectra.
The tests showed that, in most cases, the procedure successfully removed the
imprinted structure resulting from the changes in sensitivity with wavelength
that is visible in spectra calibrated with the nominal point source sensitivity
curve.

\subsection{Contamination}
\label{sec:contam}

One of the disadvantages of slitless spectroscopy is that spectra of science
targets may overlap with the spectra of random objects in the field that are
not masked by any slit mechanism. Only isolated objects are not affected by
this sort of contamination. The level of contamination depends on the
separation and relative flux level of the objects. Spectra that are
contaminated may still be useful if the expected flux from the contaminator is
significantly smaller than the flux in the spectrum of interest. Estimating the
contamination level for all spectra is essential to exploit them
scientifically.

\new{Contamination levels were estimated by aXe assuming flat contaminating
spectra \citep{axe1}. We used this feature to estimate the contamination both
from objects in the source catalogue and from bright 2MASS sources (see
Sect.~\ref{sec:extern-2mass-sourc}) that were close enough to the target object
so that their spectra might overlap.  The contamination level was estimated in
the following manner: The shape of spectra in the cross dispersion direction
was assumed to be Gaussian with a dispersion $\sigma$ equal to the object size
listed in the target catalogue.  The spectrum of each potential contaminator
was assumed to be flat in $f_{\lambda}$ with an integrated magnitude identical
to the one measured on the undispersed image.  With these assumptions,
two-dimensional spectra for all potentially contaminating objects were
produced. Finally, a spectrum at the position of the target was extracted in
exactly the same manner as the spectrum from the original grism image.  The
resulting spectrum was then used as an estimate of the contaminating flux.}

\new{
\subsection{Co-addition of Spectra}

The final step in the extraction of spectra is to sum up all flux contributions
for a given wavelength bin. The approach used by aXe is to rectify the two
dimensional spectra onto a predefined grid using the drizzle approach
\citep{drizzle}. In }\new{ our modified version of aXe, each spectrum of a
target is first drizzled to the same grid. We then use the IRAF task {\it
imcombine} with 4$\sigma$ rejection to combine the two-dimensional spectra.
One-dimensional spectra are finally produced by summing up all pixels that
correspond to the same wavelength. }

\section{The Pipeline for the Hubble Legacy Archive Grism data
(PHLAG)}\label{sec:phlag}

The whole extraction procedure from the retrieval of the input data to the
extraction of the calibrated spectra was carried out in a single Python script,
called {\bf P}ipeline for the {\bf H}ubble {\bf L}egacy {\bf A}rchive {\bf
G}rism data (PHLAG).  PHLAG works with the NICMOS calibrated images and  calls
the necessary external packages in a series of processing steps.  The steps
carried out are as follows:

\begin{enumerate}

\item {\bf Data preparation:} The data are prepared for the pipeline reduction.
The direct images are grouped according to the filter.  Pairs, consisting of
one direct image and one slitless image with small positional offsets, are
composed.

\item {\bf Image combination:} The direct images are rectified and co-added as
described in Sect.~\ref{sec:coadd}.

\item {\bf Object detection:} The object catalogues are extracted from the
undispersed image (Sect.~\ref{sec:undisp-image-catal}).

\item {\bf Spectral extraction:} The extraction of one  and two-dimensional
spectra is performed using the aXe software package with the modifications
described in Sect.~\ref{sec:extract}.

\item {\bf Metadata:} The spectra are prepared for ingestion into the database.
The metadata are collected or derived (see Sect.~\ref{sec:metadata}).
Associated products such as stamps and cutout-images are created.

\end{enumerate}

\section{Calibration of NICMOS Spectra}\label{sec:cal}

\subsection{Flatfield Cube}

The quantum efficiency of each pixel on the detector is a function of
wavelength, and this function is different for each pixel. For slitless
spectroscopic images, the usual step of correcting the quantum efficiency
variations by flatfielding has to be replaced by a wavelength  dependent
correction. Such a correction requires a three-dimensional flatfield cube with
detector coordinates and wavelength as the three axis.  

\begin{figure}
\includegraphics[bb=76 20 461 403,clip,width=0.47\hsize]{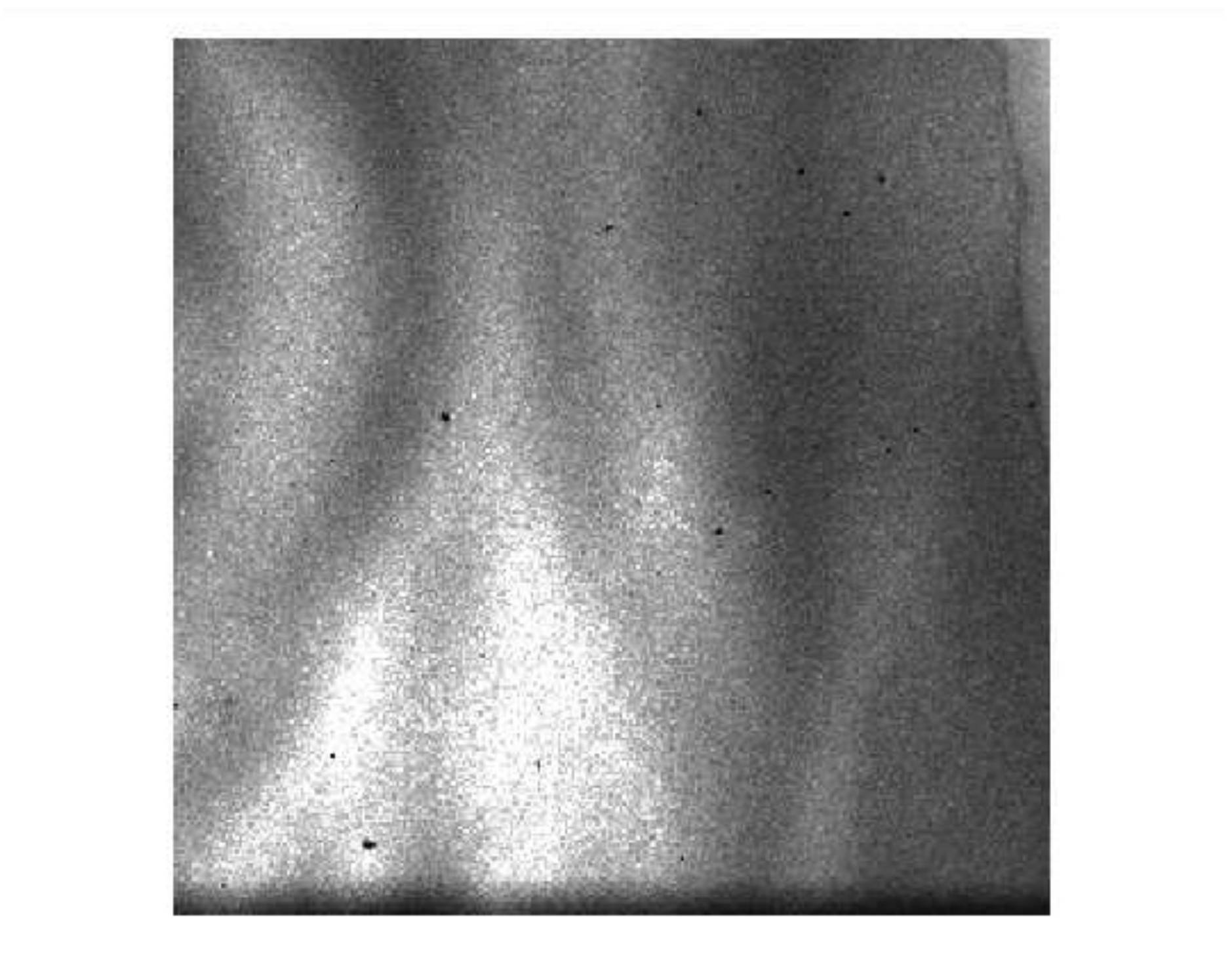}
\hfill
\includegraphics[bb=76 20 461 403,clip,width=0.47\hsize]{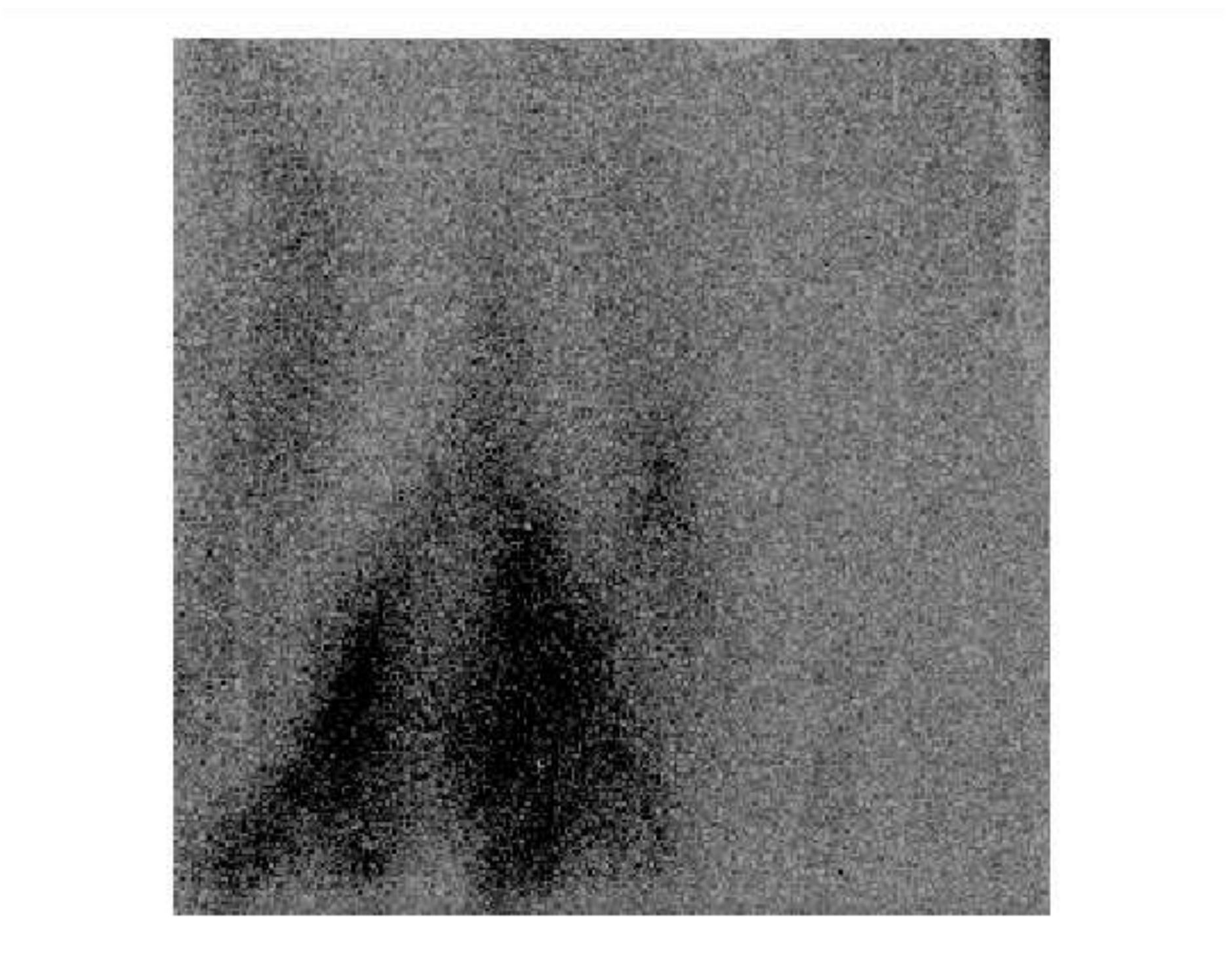}
\caption{The zeroth order (wavelength independent) plane of the 
post-NCS flat field cube (left) and the first order (wavelength
dependent) plane (right).  }
\label{Fig_flat}
\end{figure}

The flatfield cube was derived from the NICMOS narrow band flat fields,
separately for the pre-NCS and post-NCS periods.  Tab. \ref{Table_flat} lists
the narrow band flats that were used.  For each pixel, we fitted a 5th order
polynomial  to the value of the pixel in each flatfield.  Fig.~\ref{Fig_flat}
illustrates the zeroth order (wavelength independent) and first order
(wavelength dependent) terms of the flat field cube. The mean value of the rms
on the polynomial fit for the post-NCS flat field cube is 0.15\%.

\begin{table*}[t]
\begin{center}
\caption[]{Flat field cube fitting for Pre- and Post-NCS flat fields.}
\vskip 0.1cm
\label{Table_flat}
\begin{tabular}{lll} 
\hline
Filter & Pre-NCS Flat & Post-NCS Flat \\
\hline
F108N & i191346kn\_flt.fits & 3\_F108N\_STEP16\_1On-AllOff\_sflt.fits \\
F113N & i191346mn\_flt.fits & 3\_F113N\_STEP16\_1On-AllOff\_sflt.fits \\
F164N & i191346pn\_flt.fits & 3\_F164N\_STEP16\_1On-AllOff\_sflt.fits \\
F166N & i191346qn\_flt.fits & 3\_F166N\_STEP16\_1On-AllOff\_sflt.fits \\
F187N & i191346sn\_flt.fits & 3\_F187N\_STEP16\_1On-AllOff\_sflt.fits \\
F190N & i191346tn\_flt.fits & 3\_F190N\_STEP16\_1On-AllOff\_sflt.fits \\
F196N & i1913470n\_flt.fits & 3\_F196N\_STEP16\_1On-AllOff\_sflt.fits \\
F200N & i1913471n\_flt.fits & 3\_F200N\_STEP16\_1On-AllOff\_sflt.fits \\
F212N & i1913472n\_flt.fits & 3\_F212N\_STEP16\_1On-AllOff\_sflt.fits \\
F215N & i1913473n\_flt.fits & 3\_F215N\_STEP16\_1On-AllOff\_sflt.fits \\
F240M & i1913475n\_flt.fits & 3\_F240M\_STEP2\_1On-AllOff\_sflt.fits  \\
\hline
\end{tabular}
\end{center}
\end{table*}
 
\subsection{Wavelength Calibration}

\begin{figure}
\includegraphics[width=\hsize]{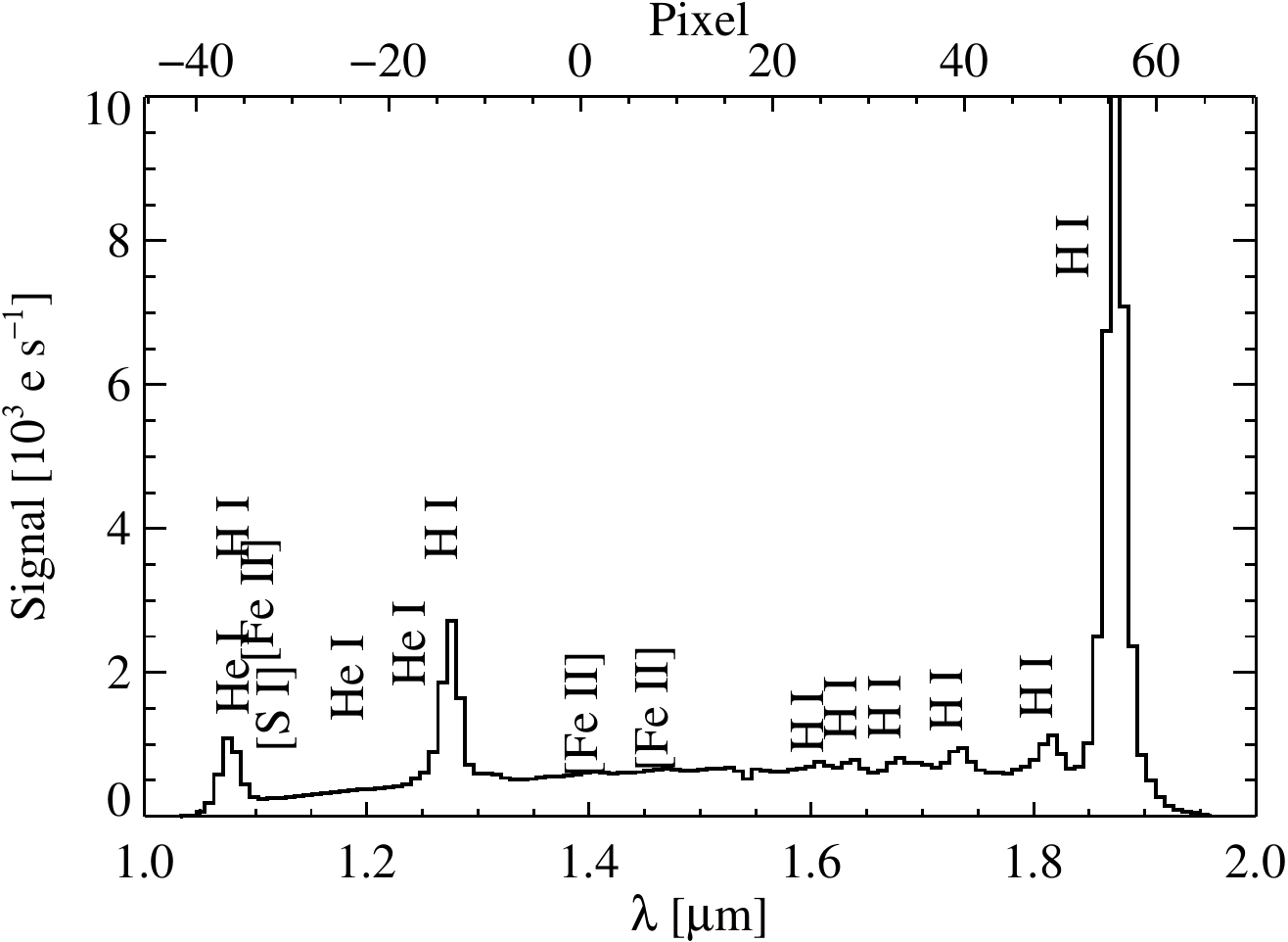}
\caption{NICMOS G141 spectrum of the planetary nebula Hb~12
showing the emission line identifications. The upper axis
shows the offset in pixels from the position of the direct image.
}
\label{Fig_Hb12}
\end{figure}

In-orbit wavelength calibration for G141 spectra was established by
observations of spectra of planetary nebulae (PNe). Two compact PNe were
observed -- Hubble 12 (Hb~12) and Vy2-2, although both are resolved by NIC3.
Fig.~\ref{Fig_Hb12} shows the one-dimensional extracted spectrum of Hb~12 with
the identified emission lines indicated. By matching the observed positions of
the lines, fitted by Gaussians, with known wavelengths, the G141 dispersion
solution was established. Vy2-2 was observed in 1997 as part of the early
calibration programme and Hb~12 in 2002 in the post-NCS era. For Hb~12 fifteen
spectra were analysed for first order dispersion solution and eight for the
second order spectrum; for Vy2-2 only three spectra were available to analyse,
and were only used to verify the solution derived from Hb~12. Within the errors
no difference was found between the wavelength solutions for both targets,
indicating no measurable change of the dispersion pre- and post-NCS. 

Up to fifteen emission lines were detectable in the first order spectra and
seven for the second order; the lines and their identifications are listed in
Tab.~\ref{Table_lines}. First and second order polynomial fits were made to the
variations of the pixel position with wavelength. The second order terms are
very small, typically $2 \times 10^{4}$ times smaller than the first order fit
coefficient; they were neglected and only a linear solution adopted. The
sampling of any spatial dependence of the wavelength solutions was poor and
single average values for the fit coefficients were adopted for the whole
detector. Tab.~\ref{Table_disps} lists the first and second spectral order
dispersion solutions in the form $ \lambda ({\rm \AA}) = A_0 + A_1  x_0$, where
$x_0$ is the pixel offset along the trace from the position of the direct
object.
 
\begin{table}
\caption[]{Nebular emission lines used for wavelength calibration.}
\label{Table_lines}
\centering 
\begin{tabular}{ll} 
\hline
~~~$\lambda({\rm\AA})$ & Species \\
\hline
10830.290 & HeI \\
10938.095 & HI \\
11164.403 & [Fe II] \\
11305.854 & [SI] \\
11969.059 & HeI \\
12527.506 & HeI \\
12818.08  & HI \\
14072.70  & [Fe II] \\
14706.272 & [Fe II] \\
16109.313 & HI \\
16407.192 & HI \\
16806.520 & HI \\
17362.108 & HI \\
18174.121 & HI \\
18751.01  & HI \\
\hline
\end{tabular}
\end{table}

\begin{table}
\caption[]{Dispersion solutions for G141 spectra.}
\vskip0.1cm
\label{Table_disps}
\centering 
\begin{tabular}{lrl} 
\hline
Spectrum & $A_0 ({\rm\AA})$ & $A_1 ({\rm\AA/pix})$ \\
order    &             &  \\
\hline
 1 & 13962.8 & 80.19 \\
 2 &  7141.7 & 40.54 \\
\hline
\end{tabular}
\end{table}

\subsection{Sensitivity Calibration}
\label{sec:fluxcal}

The sensitivity curve needed for the flux calibration of the spectra
(Sect.~\ref{sec:flux}) was derived from observations of the flux standards
GD153, G191B2B, and P330E.  Tab.~\ref{Table_fluxstd} lists the observations
used to derive the final sensitivity calibration.  Spectra were extracted from
those observations and co-added for each star separately for pre-NCS and
post-NCS observations. The sensitivity curve was then computed by dividing the
count rates of these spectra  by  tabulated standard fluxes for each star. The
standard fluxes were taken from the  HST CALSPEC library and are based on
Hubeny NLTE  models \citep{fluxcala}  in the case of G191B2B and GD153
(g191b2b\_mod\_004.fits,  gd153\_mod\_004.fits) and a combined STIS, NICMOS
spectrum in the case of P330E \citep[p330e\_stisnic\_001.fits,][]{fluxcalb}.

\begin{figure}[h]
\includegraphics[width=\hsize]{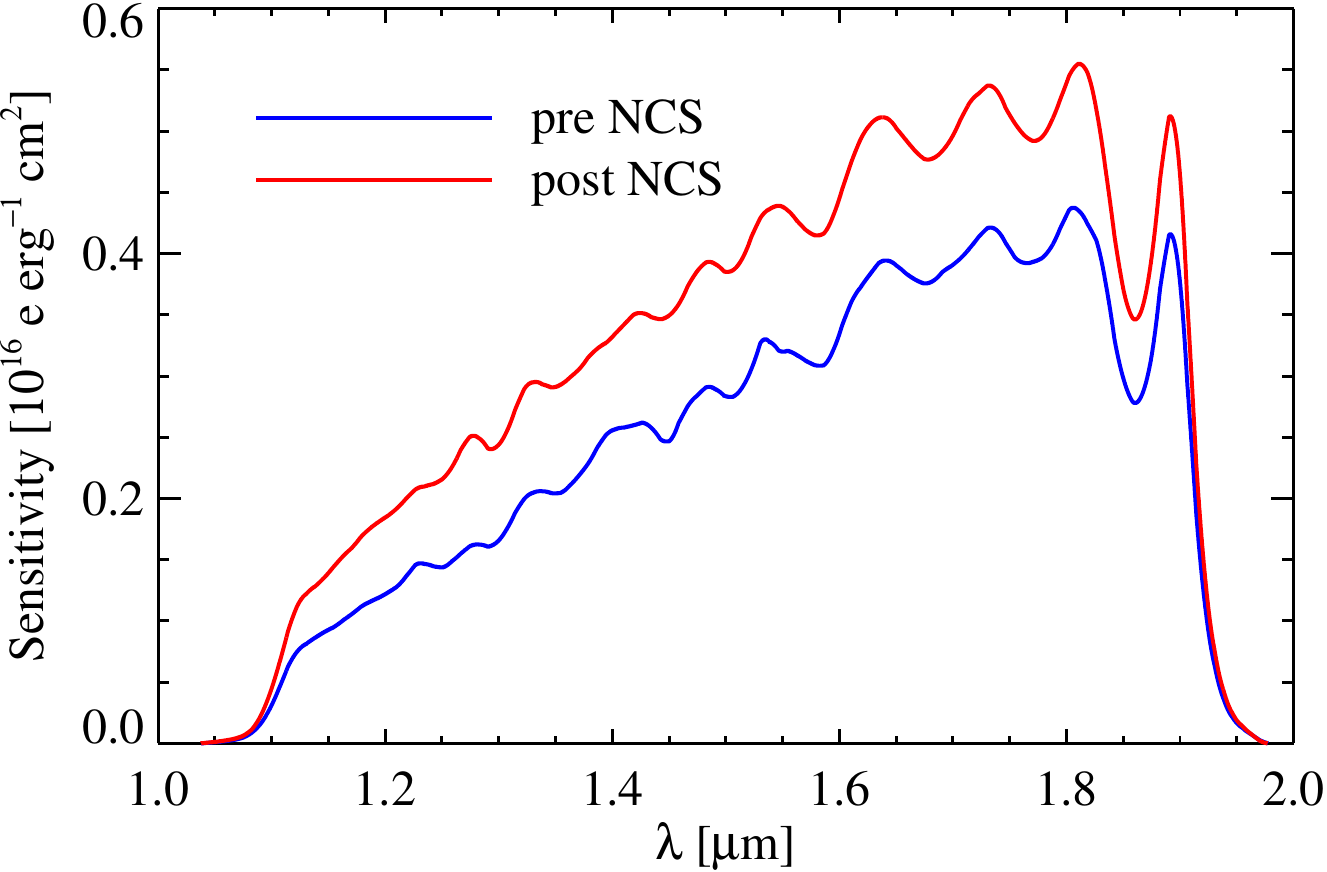}
\caption{NICMOS sensitivity curves as a function of wavelength for the
  G141 grism for pre-NCS and post-NCS data sets.}\label{fig:throughput}
\end{figure}

The average flux calibrations agreed to about 3\% between the three standard
stars and we did not detect significant variations as a function of time of
observation. However, variations as a function of  position within the field of
view indicated a scatter of about 8\%. The field of view coverage with  flux
standard stars was not sufficient to allow a field-dependent  sensitivity
calibration. Overall the absolute flux calibration for post-NCS observations
was expected to be accurate to better than 10\%, while  for pre-NCS data it was
difficult to assess the overall reliability due  to a lack of suitable standard
star observations. The sensitivity files used in the pipeline reduction carry a
random error of only 1-3\% that was propagated into the final spectra.  In
addition, there was an uncertainty of about $\sim$10\% in the overall
normalisation of the flux.  The final sensitivity curves derived for pre- and
post-NCS periods  are shown in Fig.~\ref{fig:throughput}.

\begin{table*}
\caption[]{List of flux standard stars}
\vskip0.1cm
\label{Table_fluxstd}
\centering
\begin{tabular}{llrll}
\hline
Star    & Association & PROP ID & PROP PI & pre/post NCS \\
\hline
GD153   & N94A02E6Q & 10385 & Bohlin   & post NCS \\
GD153   & N8U402MVQ &  9998 & Bohlin   & post NCS \\
GD153   & N9U203NOQ & 11064 & Bohlin   & post NCS \\
G191B2B & N9U201M3Q & 11064 & Bohlin   & post NCS \\
G191B2B & N8U405TWQ &  9998 & Bohlin   & post NCS \\
G191B2B & N94A03GDQ & 10385 & Bohlin   & post NCS \\
P330E   & N8U406VZQ &  9998 & Bohlin   & post NCS \\
P330E   & N8BR01ICQ &  8991 & Thompson & post NCS \\
P330E   & N9U212P8Q & 11064 & Bohlin   & post NCS \\
\hline
G191B2B & N4IT01NUQ &  7696 & Calzetti & pre NCS  \\
P330E   & N4VD01OYQ &  7959 & Calzetti & pre NCS  \\
\hline
\end{tabular}
\end{table*}

\section{Data Products and Distribution}

\subsection{HLA Portal}

HLA data are distributed by both ST-ECF and STScI.  There are three main ways
to search, browse, and access the NICMOS spectra:

\begin{enumerate}

\item{Archive Query Interface.} The HLA archive can be searched
online (\url{http://archive.eso.org/wdb/wdb/hla/product_science/form}) and
constraints on the target (e.g. the target name), the data properties (e.g.
effective exposure time), the source properties (e.g. the magnitude) and the
data quality (e.g. the signal-to-noise ratio) can be placed. The detailed
result pages show the preview of the data as well as all available metadata.

\item{HLA archive at STScI.} The ST-ECF HLA data has been also integrated into
the HLA interface of the Space Telescope Science
Institute (\url{http://hla.stsci.edu}). A subset of the parameters of the
ST-ECF HLA interface are shown and can be queried.

\item{Virtual Observatory.} We provide fully automated access to the HLA
metadata and data via Virtual Observatory (VO) standards. A Simple Spectrum
Access Protocol (SSAP) server has been established
(\url{http://www.stecf.org/hla-vo}).  It serves VOTables in V1.1 format, which
contain, in addition to the standard metadata, information about the footprints
of the equivalent slits of the grism spectra. Our SSAP server has been tested
with ESO's archive browser VirGO (\url{http://archive.eso.org/cms/virgo/}) as
well as with SPLAT (\url{http://star-www.dur.ac.uk/~pdraper/splat/splat-vo/}, 
also available at \url{http://starlink.jach.hawaii.edu/})
and VOSpec (\url{http://esavo.esa.int/vospecapp}). 

\end{enumerate}

\subsection{Distributed Files} \label{sec:metadata}

The spectra are distributed as sets of FITS files, which include two
dimensional cutouts of each target from the rectified and calibrated  grism
images, a cutout of the target from the undispersed filter image and  a
one-dimensional extracted spectrum.

The two-dimensional grism stamp images and the direct image cutouts are
multi-extension FITS files.  The one-dimensional spectrum follows the data
formatting specified for FITS serialisation by the IVOA Spectral Data Model
version 1.01 \citep{mcd}.  Each data point contains the wavelength in \AA, the
count rate in electrons per second and the flux expressed in physical units
along with associated errors and an estimate of the contaminating flux.
Additional metadata include keywords to describe the contamination, the
orientation of the dispersion direction on the sky and footprints.

\begin{figure}
\includegraphics[width=\hsize]{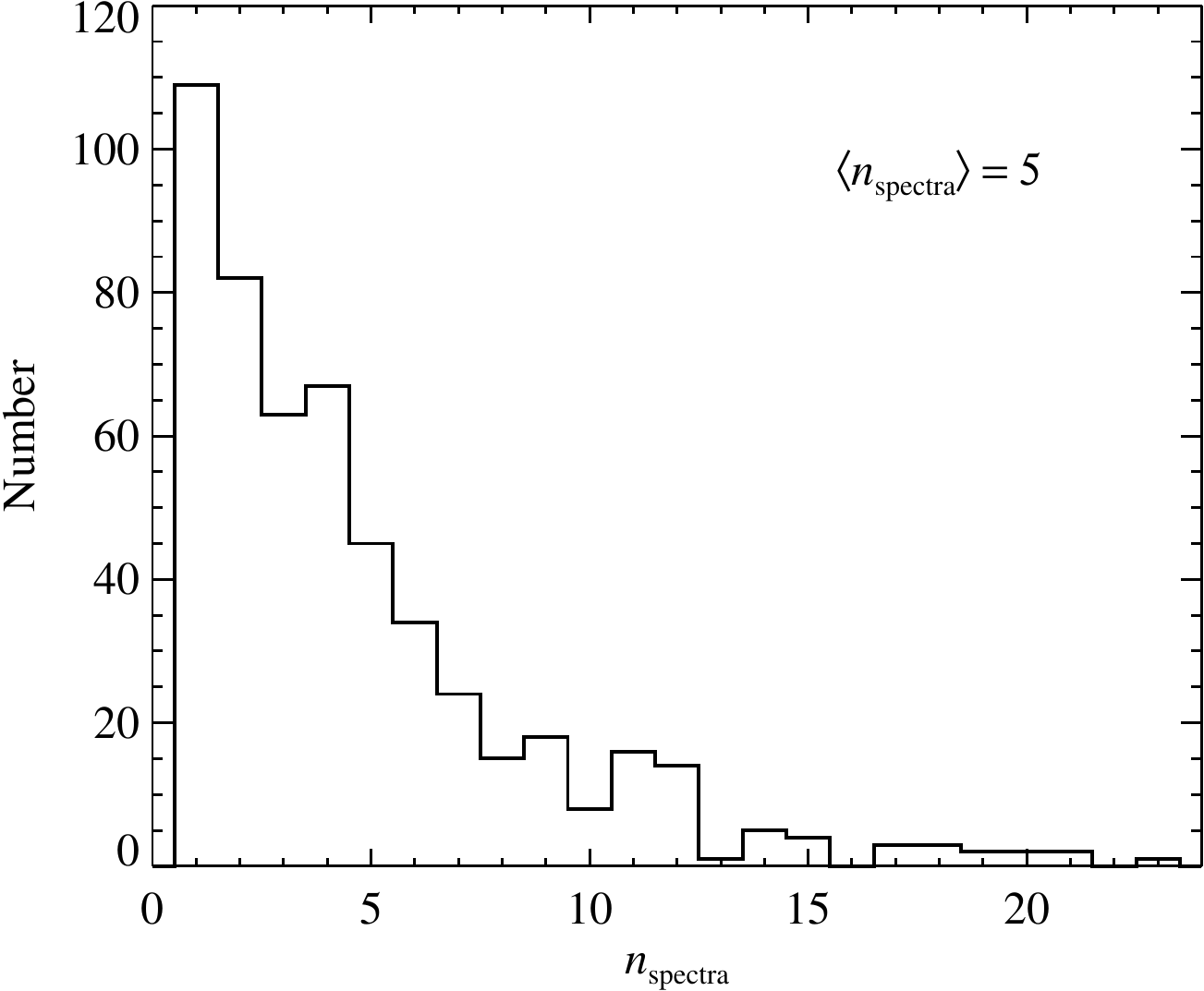}
\caption{Distribution of the number of extracted spectra from HLA
  datasets.}\label{fig:datasets}
\end{figure}

\begin{figure}
\includegraphics[width=\hsize]{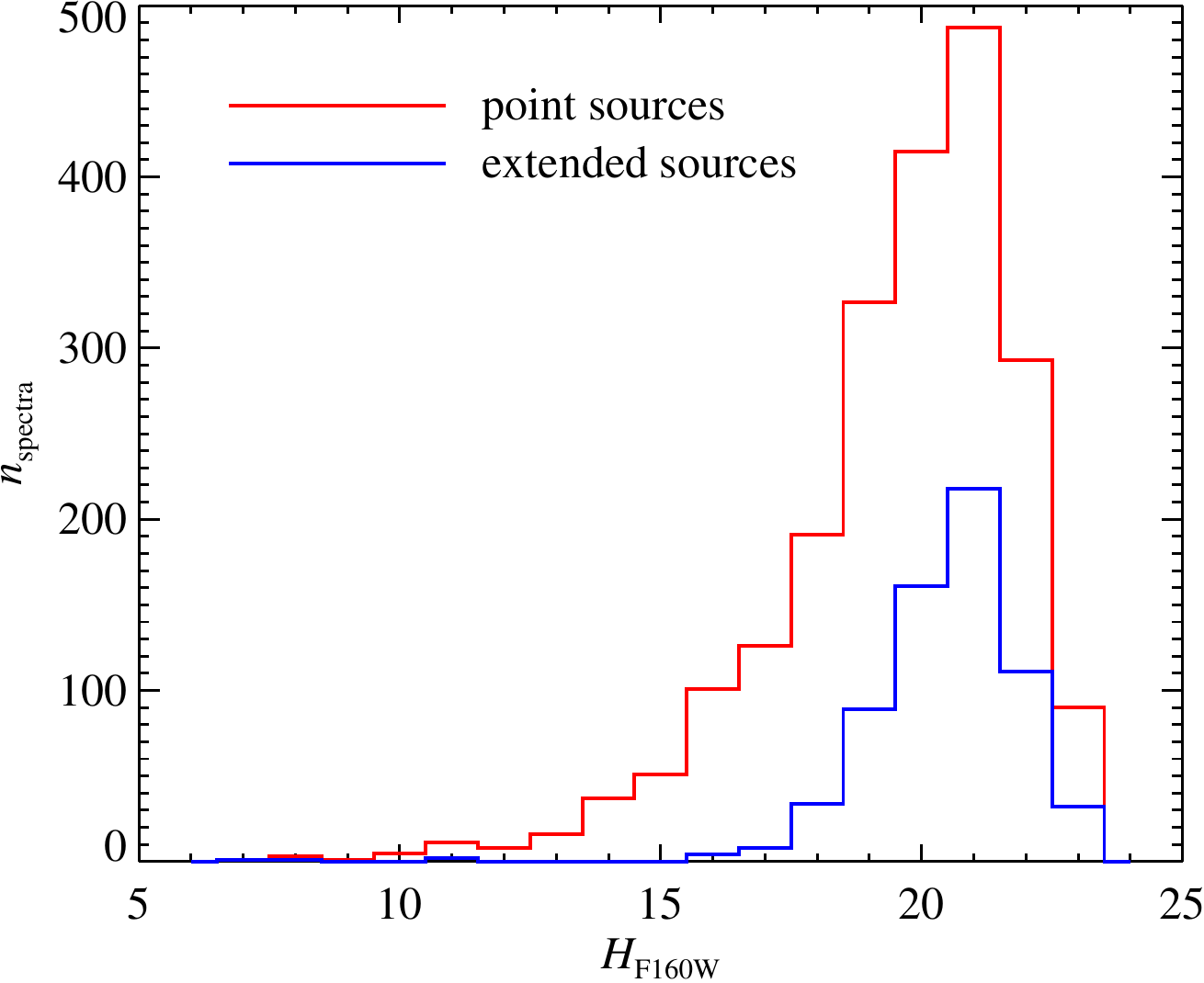}
\caption{Magnitude distribution of the targets with spectra in the HLA release. The blue
histogram are the extended sources. }\label{fig:maghist}
\end{figure}

\begin{figure}
\includegraphics[width=\hsize]{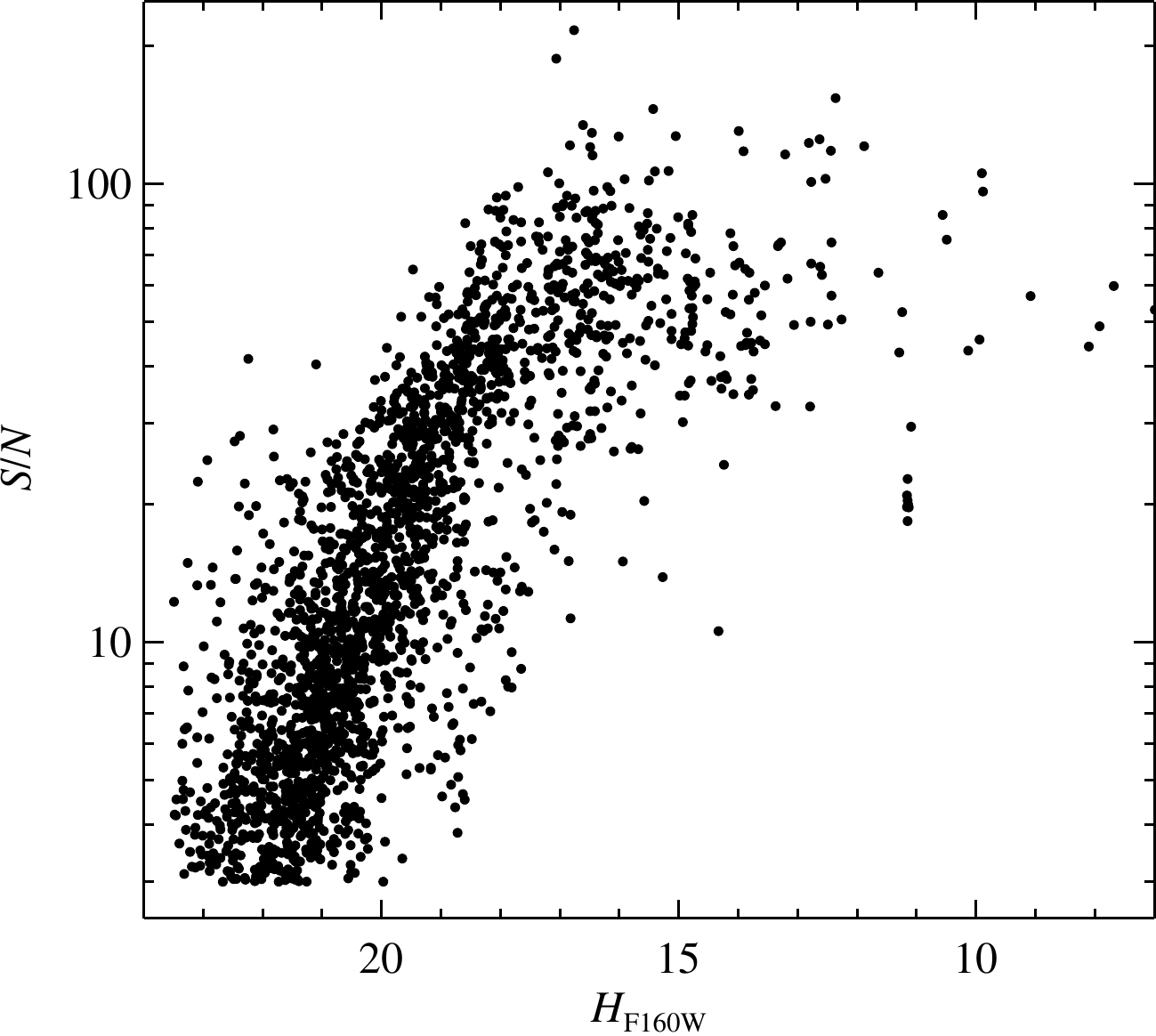}
\caption{Measured signal-to-noise of the HLA spectra as a function of magnitude.}\label{fig:snmags}
\end{figure}

\section{Results}\label{sec:qc}

\subsection{Data Release 1}

Data were released on \releasedate. This Data Release 1 (DR1) includes a total
of \nspectra\ extracted from  \nassociations\ associations. 

The H-band magnitudes computed from the undispersed images of the targets range
from 7 to 23.5  for point sources, and from 16 to 23.5 for extended sources as
identified by the SExtractor program (see Fig.~\ref{fig:maghist}).  Only
spectra with measured signal-to-noise ratios larger than three were included in
the release, and  typical measured signal-to-noise ratios are about 50 for
objects brighter than H$\approx$17 (see Fig.~\ref{fig:snmags}).

\subsection{Completeness}

Whether the spectrum of a particular target is extracted depends, among other
things, on the detection of the target on the undispersed image, the  relative
location and flux levels of other spectra on the grism image, the location of
residual image defects and  the location of the spectrum  relative to the image
edges.  The number of spectra extracted at each pointing therefore varies
widely, and was typically about five.  The distribution of the number of
spectra extracted from each association is shown in Fig.~\ref{fig:datasets}.

\new{To assess the completeness of the sample quantitatively, we have compared
the DR1 catalogue with the survey by \citet{pat}. These authors searched
two-dimensional NICMOS grism images for emission line objects and found a total
of 33 candidates. Most of the emission lines are H$\alpha$ at redshifts between
0.75 and 1.9. Such a sample of galaxies is dominated by closely interacting
systems and, because of contamination issues, our procedure selects against
such objects. We found that a total of 6 of the \citeauthor{pat} candidates are
included in DR1. We therefore consider 20\% to be a lower limit for the
completeness of our catalogue.  }

The HLA includes any spectrum suitable for extraction, no attempt has been made
to classify or match the targets with any catalogue. Some of the fields  have
been observed several times with different roll angle. These spectra were not
co-added and therefore several spectra might be available for one target.  In
total, the number of unique targets in the release is \ntargets.

\subsection{Photometric accuracy of the undispersed images}
\label{sec:phot-accur-undisp}

The photometric accuracy of the undispersed direct images was assessed by
comparing the magnitudes obtained in the F160W and F110W bands with the 2MASS H
and J band magnitudes of matched objects.  When doing this comparison, we used
aperture photometry with an aperture of 11 pixels, and applied a
finite-aperture correction as provided by the NICMOS Data Handbook
\citep{nicmoshandbook}; in addition, we converted the 2MASS magnitudes into the
AB magnitude system.  

\begin{figure}[h]
  \centering
  \includegraphics[width=\hsize]{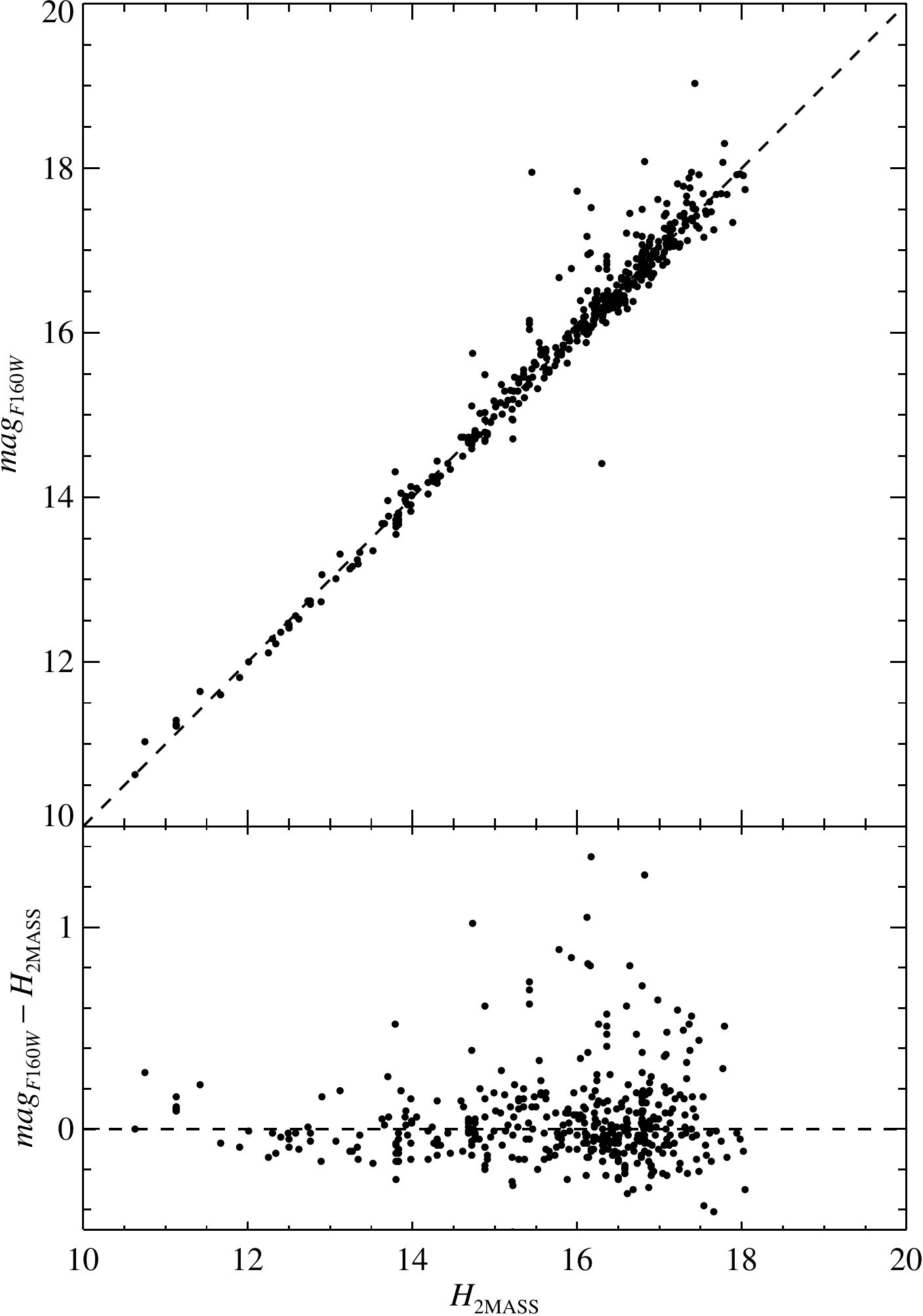}
  \caption{\textbf{Top.} Direct image F160W magnitude, as a function
    of the 2MASS $H$ magnitude.  \textbf{Bottom.}  Difference of the
    two magnitudes as a function of the 2MASS magnitude .}
  \label{fig:7}
\end{figure}

The results obtained are shown in Fig.~\ref{fig:7}: a good agreement is
obtained over the whole magnitude range, with negligible bias; the relatively
large scatter observed, $\sim0.16 \mbox{ mag}$, is probably due to the
photometric uncertainties present in the 2MASS catalogue for faint sources.
Similar results were obtained for the F110W magnitudes.

\subsection{Photometric accuracy of the spectra}

\begin{figure}[t]
  \centering
  \includegraphics[width=\hsize]{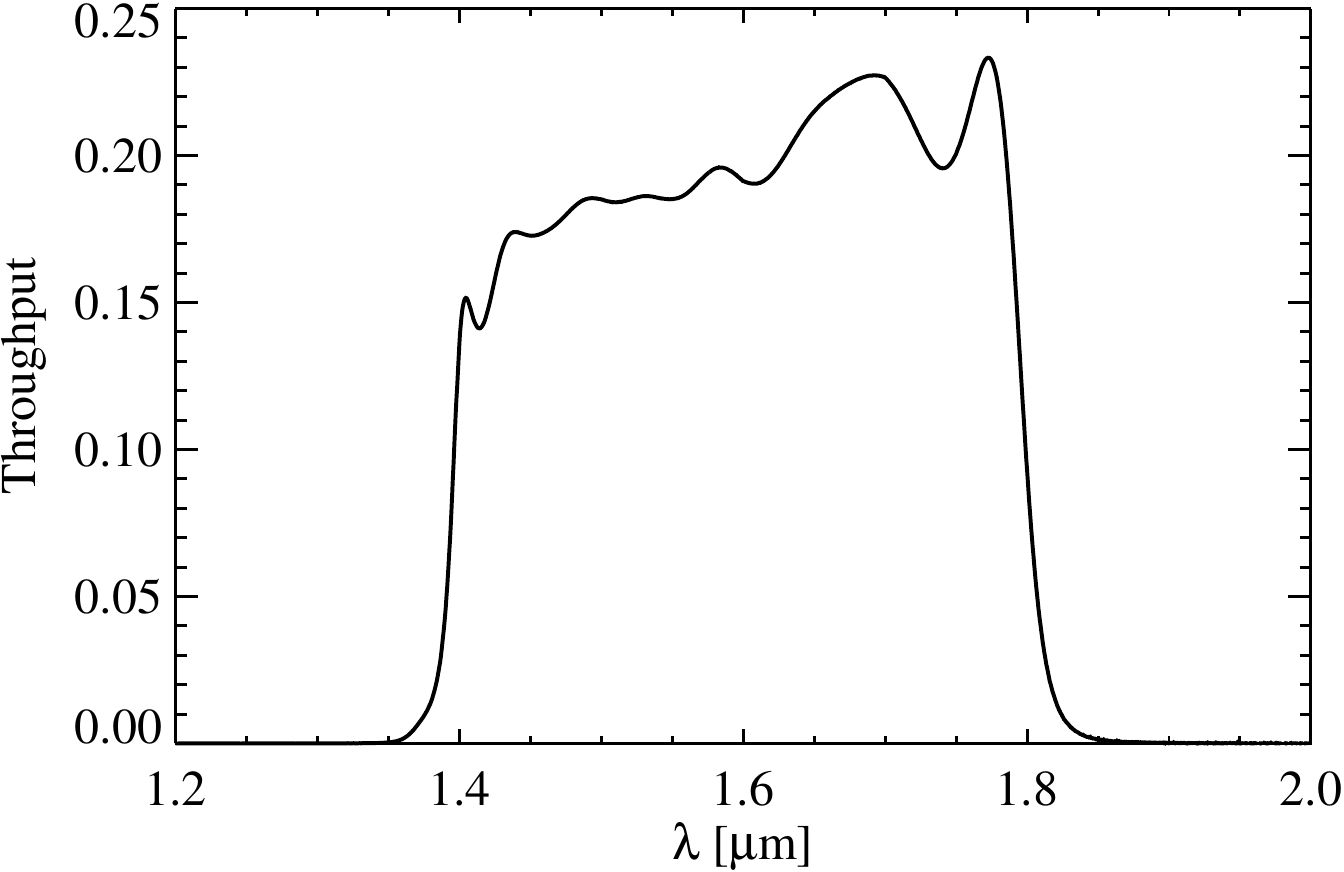}
  \caption{Total throughput of the F160W filter for the NICMOS/NIC3 camera.}
  \label{fig:4}
\end{figure}

Quality control of the photometric calibration of the one-dimensional spectra
was performed by comparing the integrated flux, as measured from the
end-product spectra, with the magnitude derived from the direct images.  The
following procedure was applied: 

\begin{itemize} 

\item all spectra  that had associated direct images in the F160W filter were
selected; 

\item for the selected objects, only spectra which were considered complete up
to the boundary of the total throughput of the F160W/NIC3 configuration (see
Fig.~\ref{fig:4}), were selected.  In particular, all spectra that did not
contain data within the wavelength range defined by $\mathrm{throughput} >
0.005$ were discarded; 

\item the flux of these objects was then integrated and compared with either
the direct image magnitudes or, when available, with the 2MASS $H$ band
magnitudes.  At the same time, the error on the magnitude due to flux
contamination from other sources was also evaluated using data provided in the
output products.  

\end{itemize}

\begin{figure}
  \centering
  \includegraphics[width=\hsize]{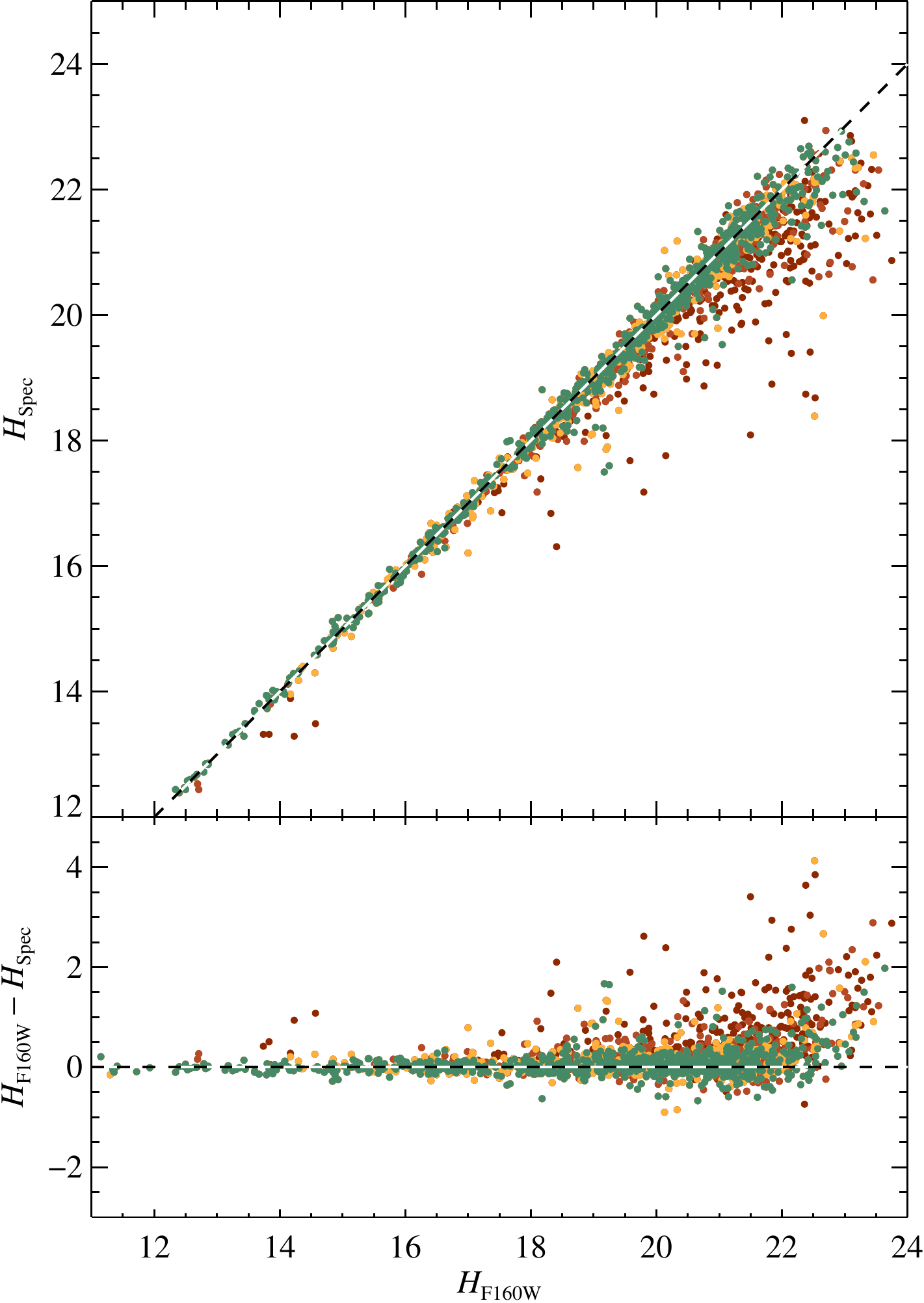}
  \caption{\textbf{Top.} The ``spectroscopic'' magnitude, i.e.\ the
    magnitude derived from the spectrum of the object, is shown as a
    function of the direct image F160W magnitude.   \textbf{Bottom.}
    The difference of the two magnitudes is plotted as a function of
    the direct image magnitude.  \new{For both plots, the colour
    of the points indicates the level of contamination, 
coded according to the difference $\Delta m$ in the magnitude  of the target
and that of the contaminator. The colour codes are as follows:
 green:              $\Delta m > 8^{\sl mag}$,
 yellow: $4^{\sl mag}<\Delta m < 8^{\sl mag}$,
 orange: $2^{\sl mag}<\Delta m < 4^{\sl mag}$
and  red: $0^{\sl mag}<\Delta m < 2^{\sl mag}$.
 }
}
  \label{fig:5}
\end{figure}

\begin{figure}[h]
  \includegraphics[height=0.32\hsize]{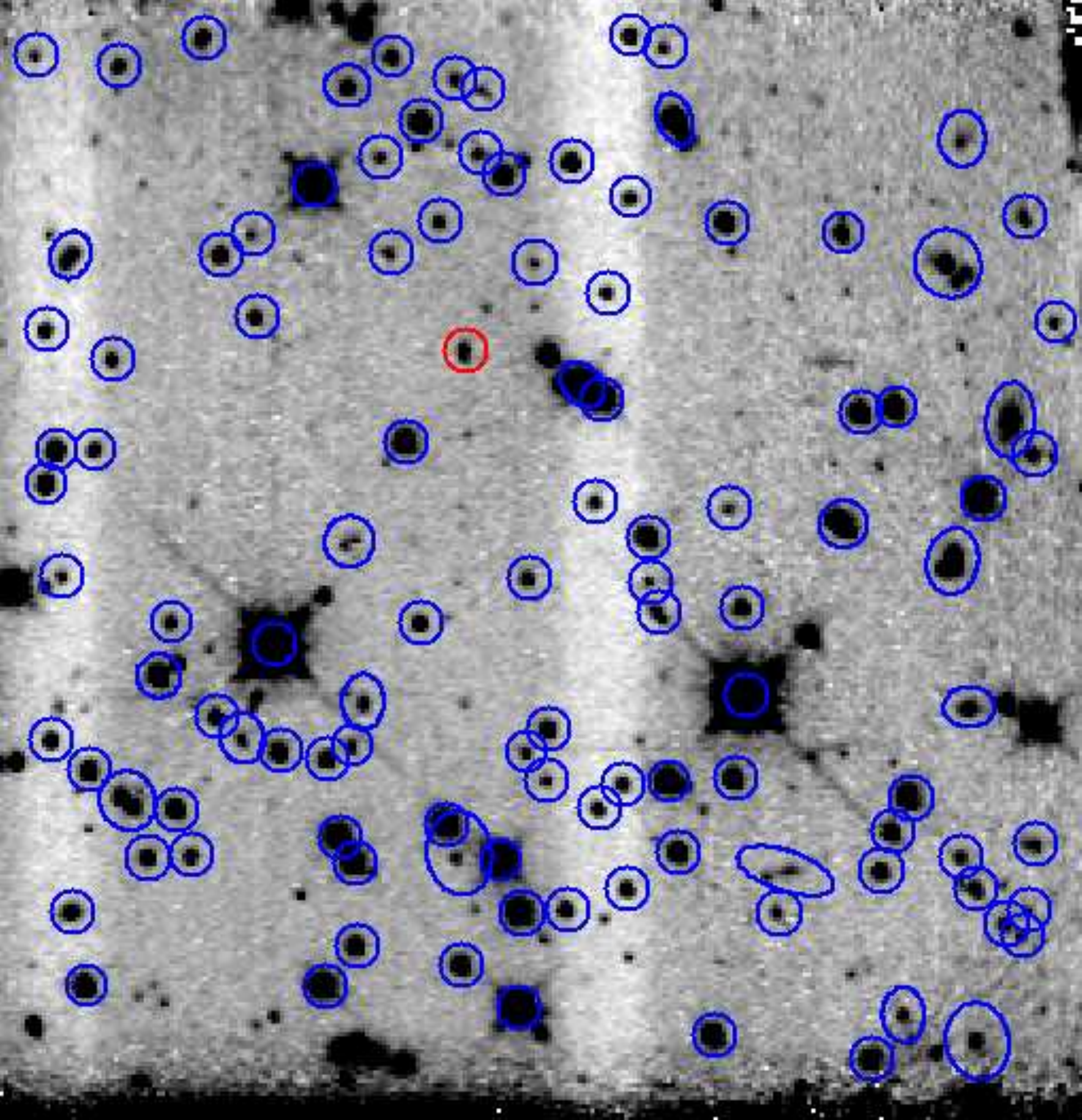}
  \hfill
  \includegraphics[height=0.32\hsize]{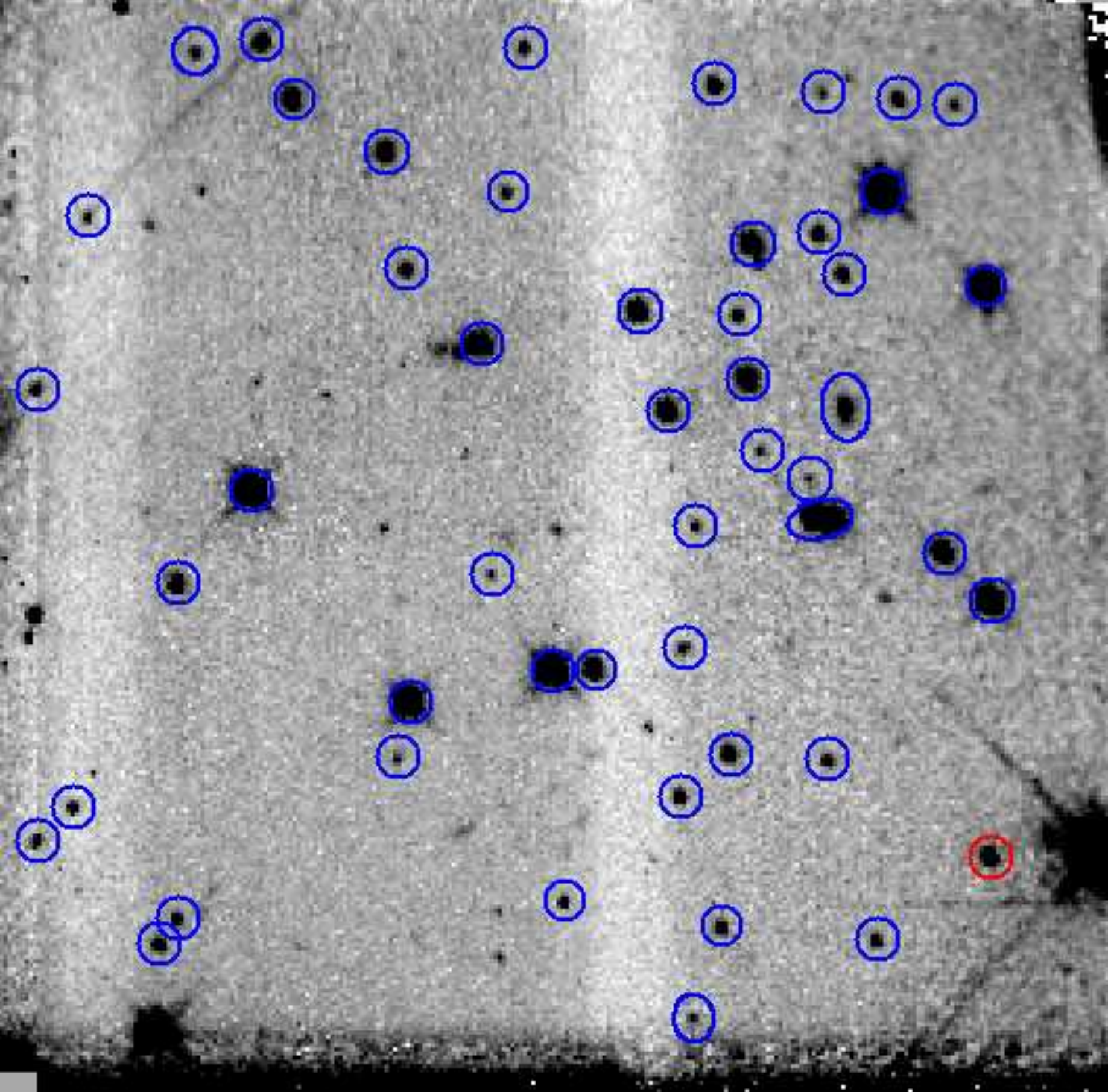}
  \hfill
  \includegraphics[height=0.32\hsize]{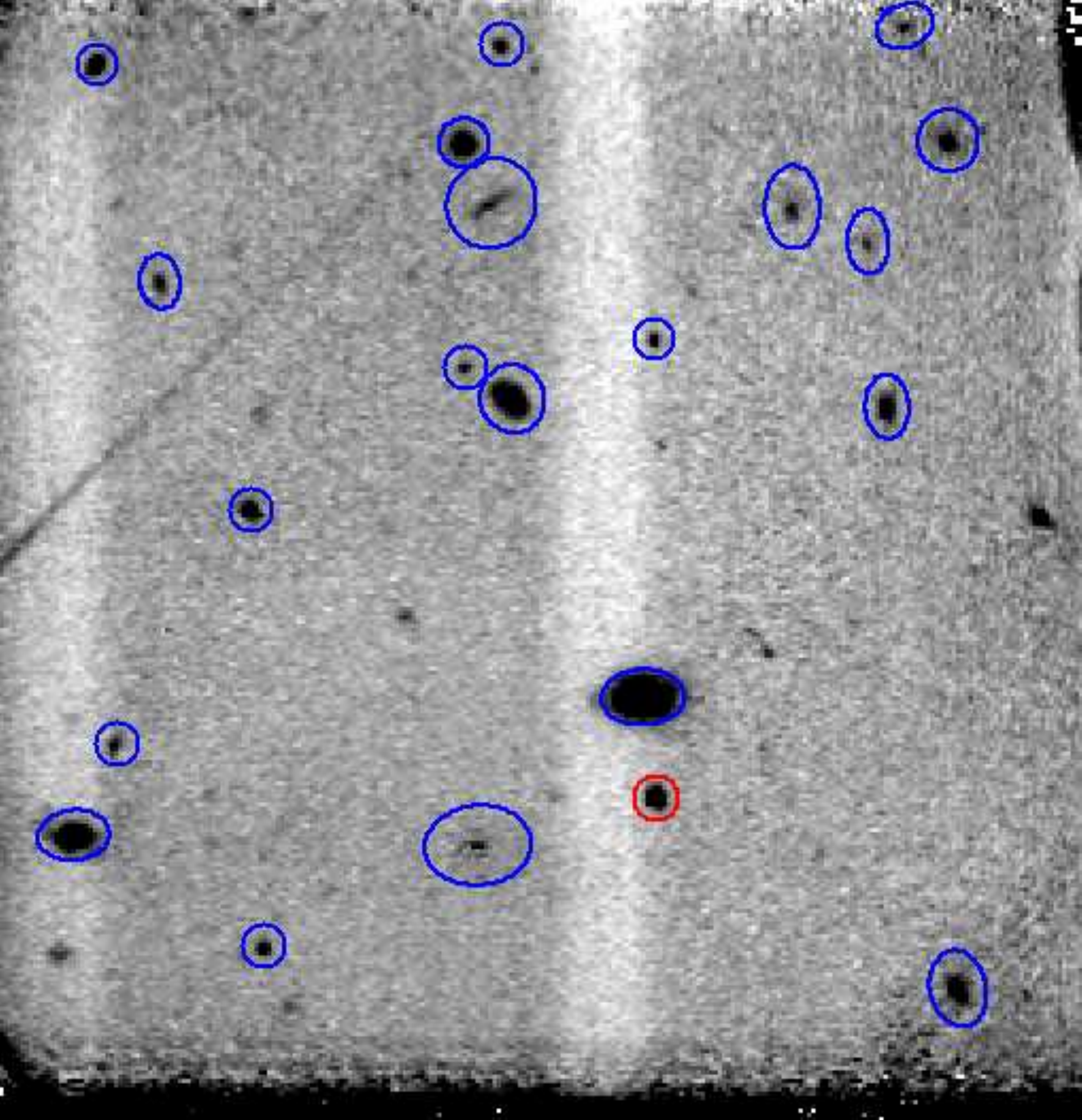}
\newline
  \includegraphics[width=0.31\hsize]{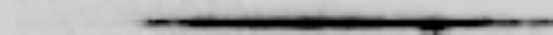}
  \hfill
  \includegraphics[width=0.32\hsize]{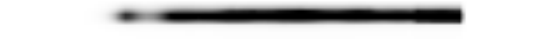}
  \hfill
  \includegraphics[width=0.31\hsize]{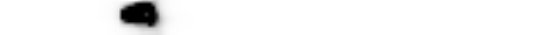}
  \caption{Examples of catastrophic discrepancies selected from the
    outliers of Fig.~\ref{fig:5}.  Each panel shows
    the direct image associated to the source for which an inaccurate
    spectroscopic magnitude was obtained (marked in red); other
    objects in the field used for the contamination calculation are
    marked with blue ellipses. \new{The corresponding cutouts from
    the grism image are shown below the undispersed images.}
    \textbf{Left.} Source in a crowded
    field, close to an undetected source to the
    right. \textbf{Middle.} Source close to bright star on the
    boundary. \textbf{Right.} Source contaminated by a bright star
    outside the field (note the spike on the centre-left of the
    image).}
  \label{fig:6}
\end{figure}

The results obtained for the comparison with the F160W direct images are shown
in Fig.~\ref{fig:5}.  A very good linear response is obtained over a large
range of magnitudes for both pre- and post-NCS data.  The effects of spectrum
contamination are visible in the lower-right part of the graph, which is
occupied mainly by \new{sources with  high levels of contamination, as directly
evaluated from the spectra}.  Overall the agreement is good, with a median
offset of 0.16 magnitudes for pre-NCS data, and 0.23 for the post-NCS case.
The scatter is in both cases $\sim0.5$ magnitudes.

A few ``catastrophic'' cases of sources with apparent no-contamination but
still large magnitude difference are however evident in this plot.  These cases
were investigated further and simple explanations could be found for almost all
of them.  In general, it was found that large discrepancies were associated
with: (i) crowded fields, for which the contamination model might be
inaccurate; (ii) objects close to bright sources not included in the analysis;
(iii) objects close to extended, diffuse sources; or (iv) objects contaminated
by sources outside the field of view of the direct image. \new{ Three examples
of such cases are shown in Fig.~\ref{fig:6}. For each case, the undispersed
image  with the marked problematic target is shown along with the corresponding
spectrum cutout. In the left and centre panels, a bright undetected source
produces significant  contamination in the extracted spectra.  In both cases,
the contaminating source dominates the flux. In the right panel, the zeroth
order of a very bright source outside the field produces a bright region  in
the extracted spectrum.}

\subsection{Noise}\label{sec:sn}

\begin{figure}[ht]
  \centering
  \includegraphics[width=\hsize]{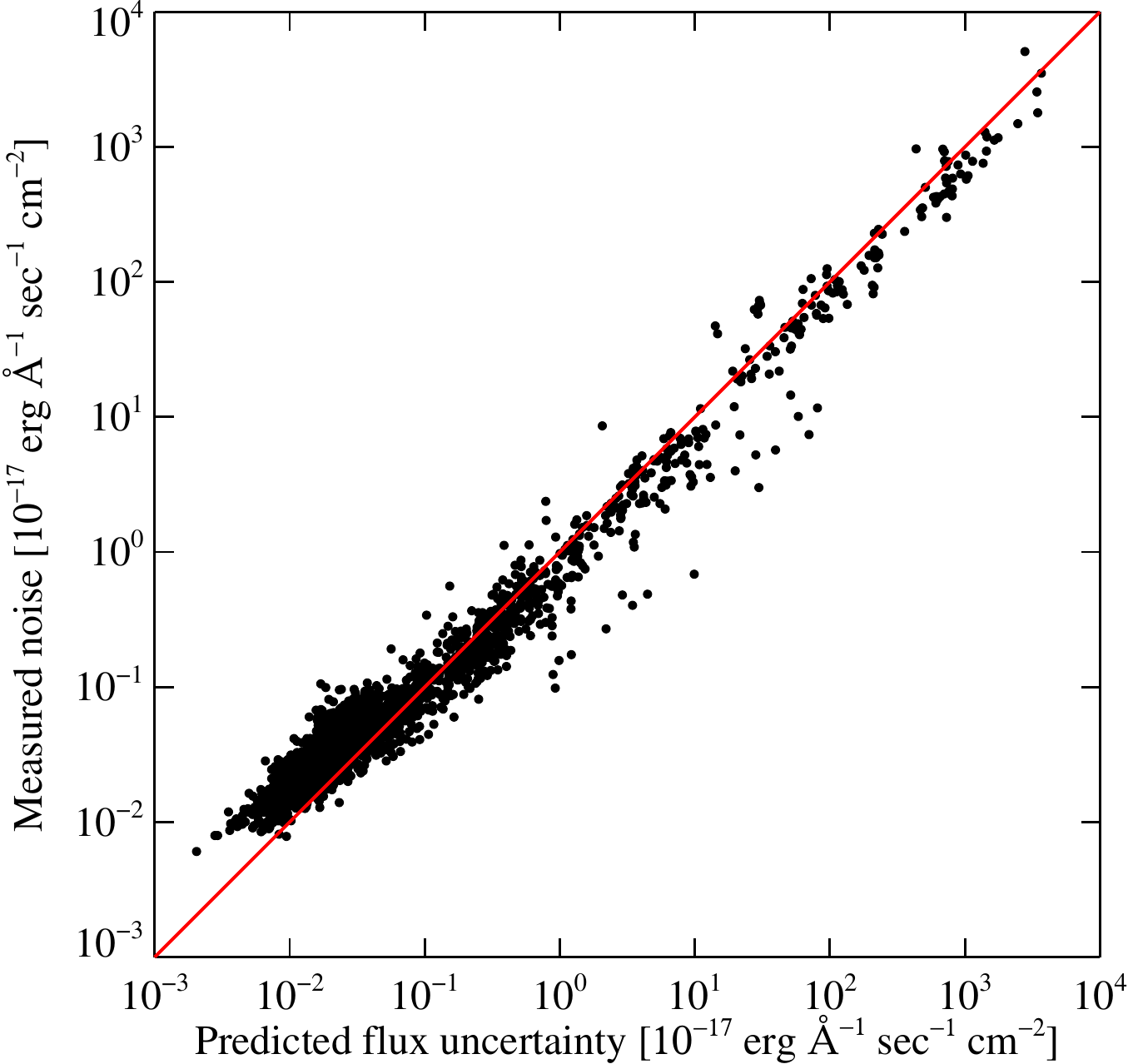}
  \caption{Noise measurements from the spectrum using the DER\_SNR
    algorithm as a function the noise estimate obtained from the
    NICMOS error array using error propagation.}
  \label{fig:signaltonoise}
\end{figure}

All extracted spectra include a predicted flux uncertainty  in each wavelength
bin based on error propagation from the error array in the NICMOS images
through to the final spectra.  Fig.~\ref{fig:signaltonoise} compares these
predicted uncertainties with noise measurements on the final spectra using the
DER\_SNR algorithm \citep{2007STECF..42....4S}. 

The SNR values are in good agreement with quite small scatter at the high flux
end. At the faint end the scatter is larger and there is a slight but
systematic trend for the DER\_SNR values to be higher than their predicted
value. This might be due to a component, e.g. the noise in the background,
which was not included in our noise model, or reflect some bias in the
measurement of the noise.

\subsection{\new{Comparison of Spectra with NICMOSlook Extractions}}

\begin{figure}
\includegraphics[width=\hsize]{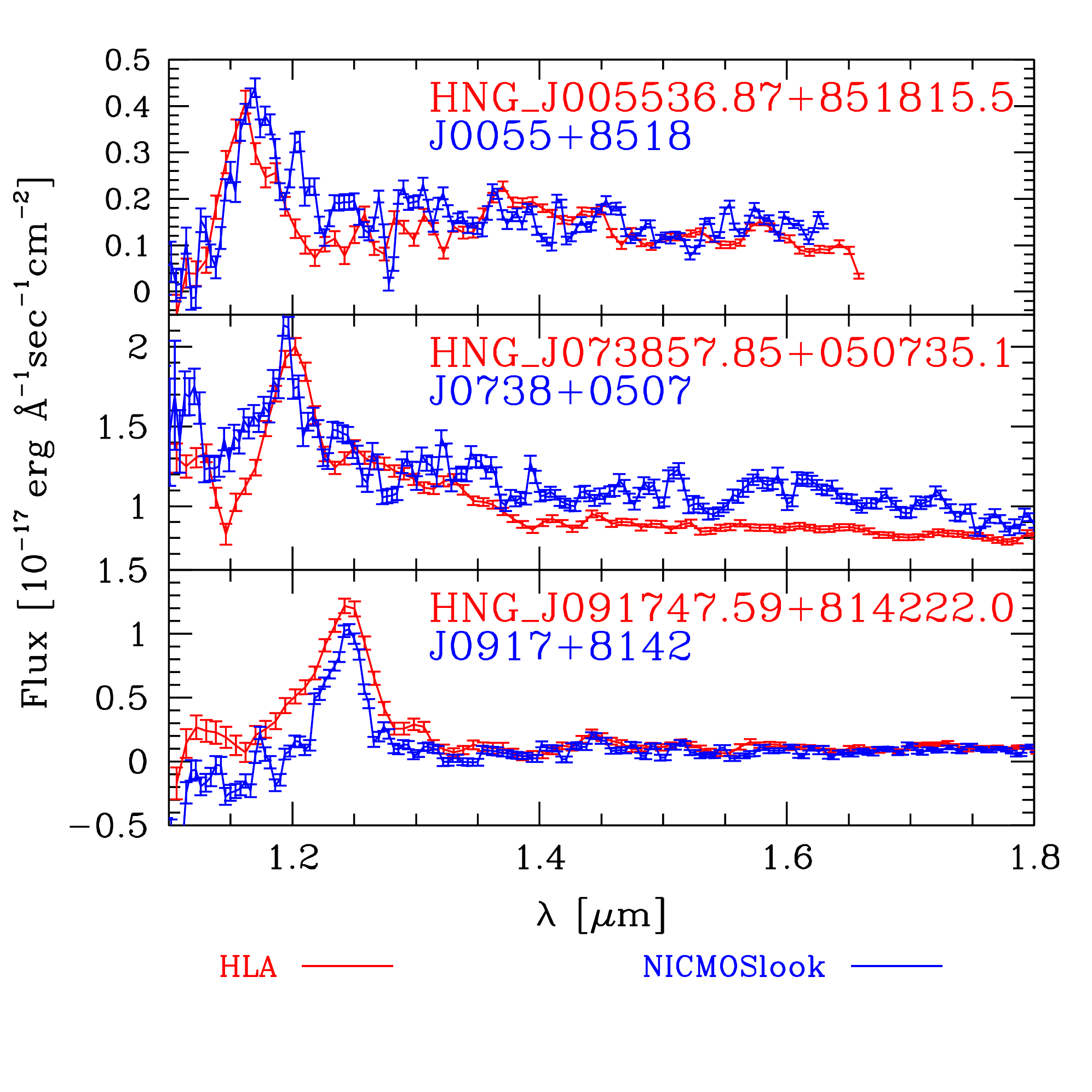}
\caption{\new{Comparison of HLA spectra  with extractions using 
the NICMOSlook program. All three spectra are H$\alpha$ emitting galaxies
 discovered by \cite{pat}. The blue curves are NICMOSlook extractions, whereas the
red curves are the HLA extractions. The blue labels are the \citeauthor{pat}'s names.
} }\label{fig:nicmoslook}
\end{figure}

\new{Most previous extractions of NICMOS grism spectra have used interactive
tools, such as the NICMOSlook program \citep{nicmoslook}. NICMOSlook is a
highly specialised tool that provides a large number of options and parameters
that can be varied to optimise the extraction.  The main advantages of using an
interactive program is that the background region can be adjusted for each
individual spectrum and that the subpixel accuracy of the trace necessary for
the pixel-response correction} \new{can be visually checked and fitting
parameters can be modified.  In addition, NICMOSlook's optional deblending of
overlapping spectra results in a more reliable estimate of contamination
levels, and is accurate enough to recover heavily contaminated spectra.  The
combination of optimising the background regions and deblending of overlapping
spectra allows NICMOSlook to extract more spectra in crowded regions than is
possible  with the automatic procedure used in this work. For example,
\cite{pat} used NICMOSlook to extract the larger sample of H$\alpha$ emitters
mentioned above.

For those spectra that are included in DR1, the main uncertainty is the
background level. The impact is most visible in low signal-to-noise spectra.
In Fig.~\ref{fig:nicmoslook}, we compare  the NICMOSlook extraction of three
spectra  of the \citeauthor{pat} targets with the one from DR1, all of them are
low signal-to-noise spectra.  The good agreement of the spectra suggests that
both the relative and absolute flux calibration in DR1 is reliable.

To summarise, the accuracy of the completely unsupervised extractions of NICMOS
spectra in HLA DR1 is close to what can be achieved with an interactive tool.
In addition, we have taken into account contamination by sources outside the
field of view of NICMOS, which is not possible using currently available
interactive tools.  }

\subsection{Comparison of Spectra with Published Data}

As a quality check, we compared some extracted spectra to spectra of the same
objects taken from the ground, as well as previously published extractions of
spectra from the same NICMOS data. Because of strong atmospheric lines, H-band
IR spectroscopy is difficult to calibrate from the ground. In particular, the
region around $\lambda=1.4-1.5\,\mu$ m is not accessible from the ground. 

\begin{figure}
\includegraphics[width=\hsize]{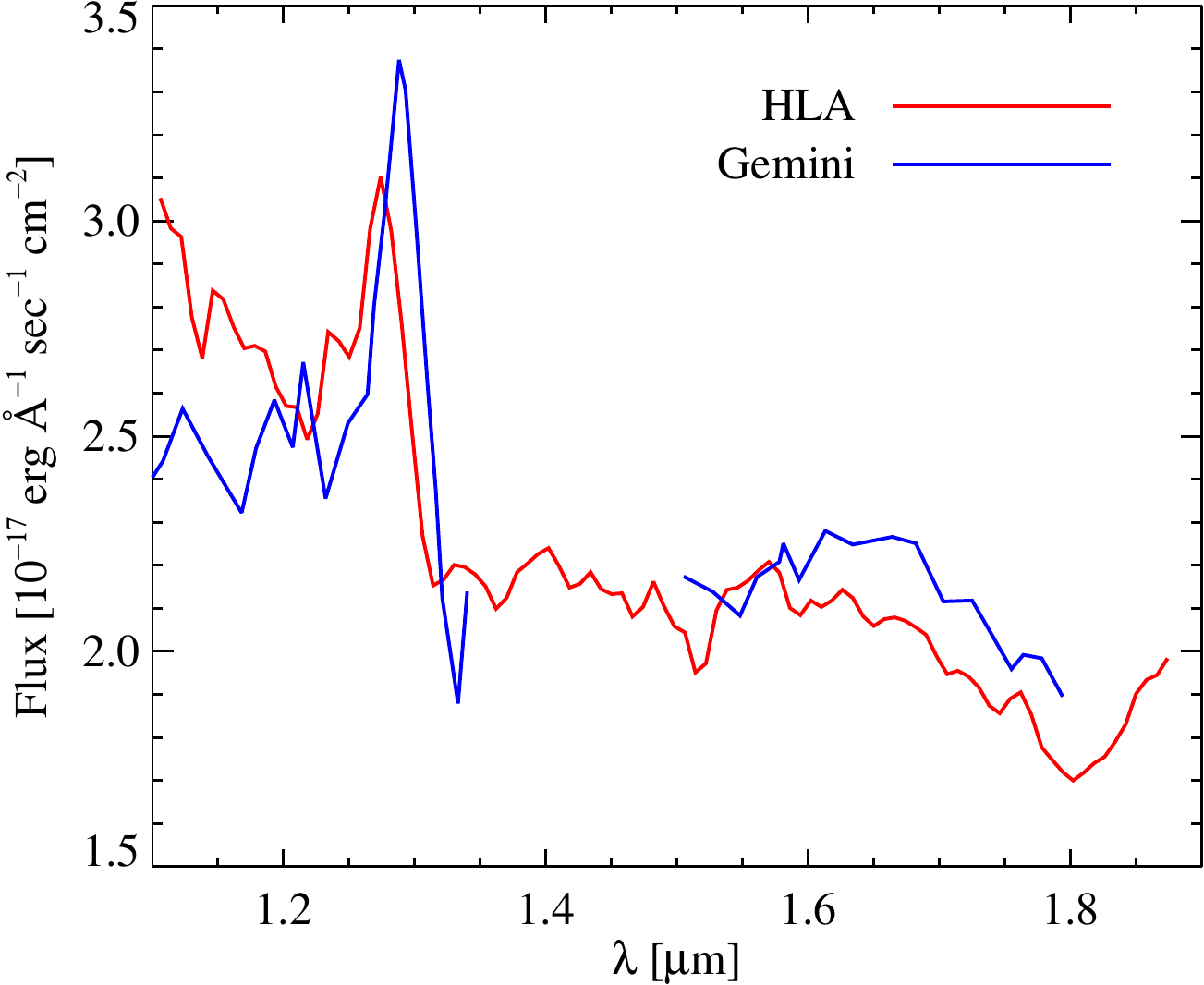}
\caption{Comparison of an HLA spectrum with a Gemini spectrum of the same
source.  The blue curve is the heavily smoothed spectrum of SDSS
J083643.85+005453.3 from \cite{comp1}, the red curve is
HNG\_J083643.82+005453.4\_N6LE01ULQ.}\label{fig:comp1}
\end{figure}

\begin{figure}
\includegraphics[width=\hsize]{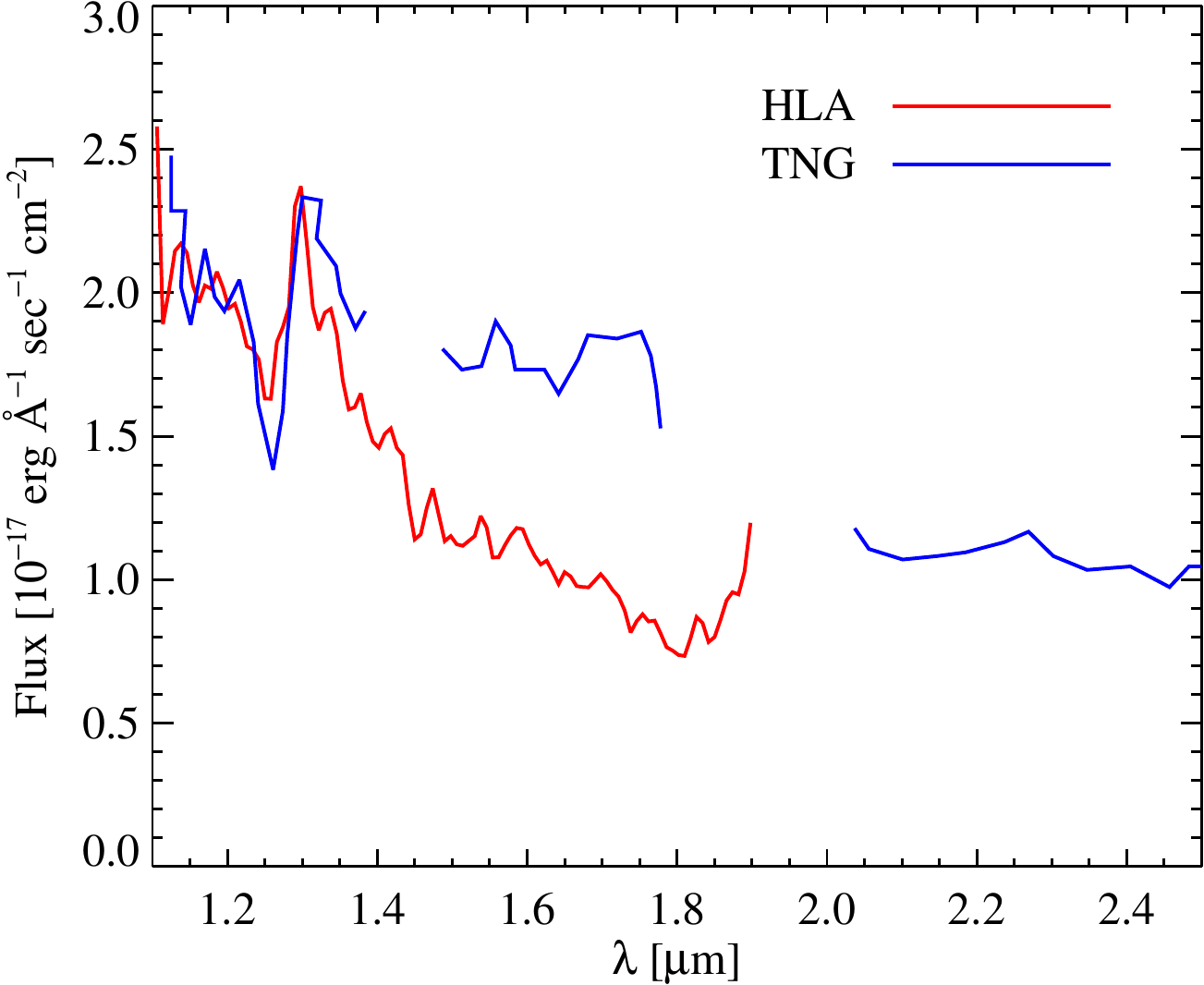}
\caption{Comparison of an HLA spectrum with a TNG NICS spectrum of the same
source.  The blue curve is the spectrum of SDSS 104433.04+012502.2 from
\cite{maio}, the red curve is
HNG\_J104433.08-012501.6\_N6LE03C7Q.}\label{fig:1044} 
\end{figure}

In Figs.~\ref{fig:comp1} and~\ref{fig:1044}, we compare spectra of two high
redshift QSOs to spectra from Gemini \citep{comp1} and the Italian Telescopio
Nazionale Galileo  \citep[TNG,][]{maio}. In both cases, the shape of the line
and continuum around the  CIII line at $\lambda\sim1.3\,\mu$m is well
reproduced. The continuum at wavelengths longer than 1.5$\,\mu$m agrees to
within about 5\% with the Gemini data, but differs by almost a factor of two
from the TNG data. We attribute this discrepancy to the uncertain calibration
of the data from the ground. 

\begin{figure}
\includegraphics[width=\hsize]{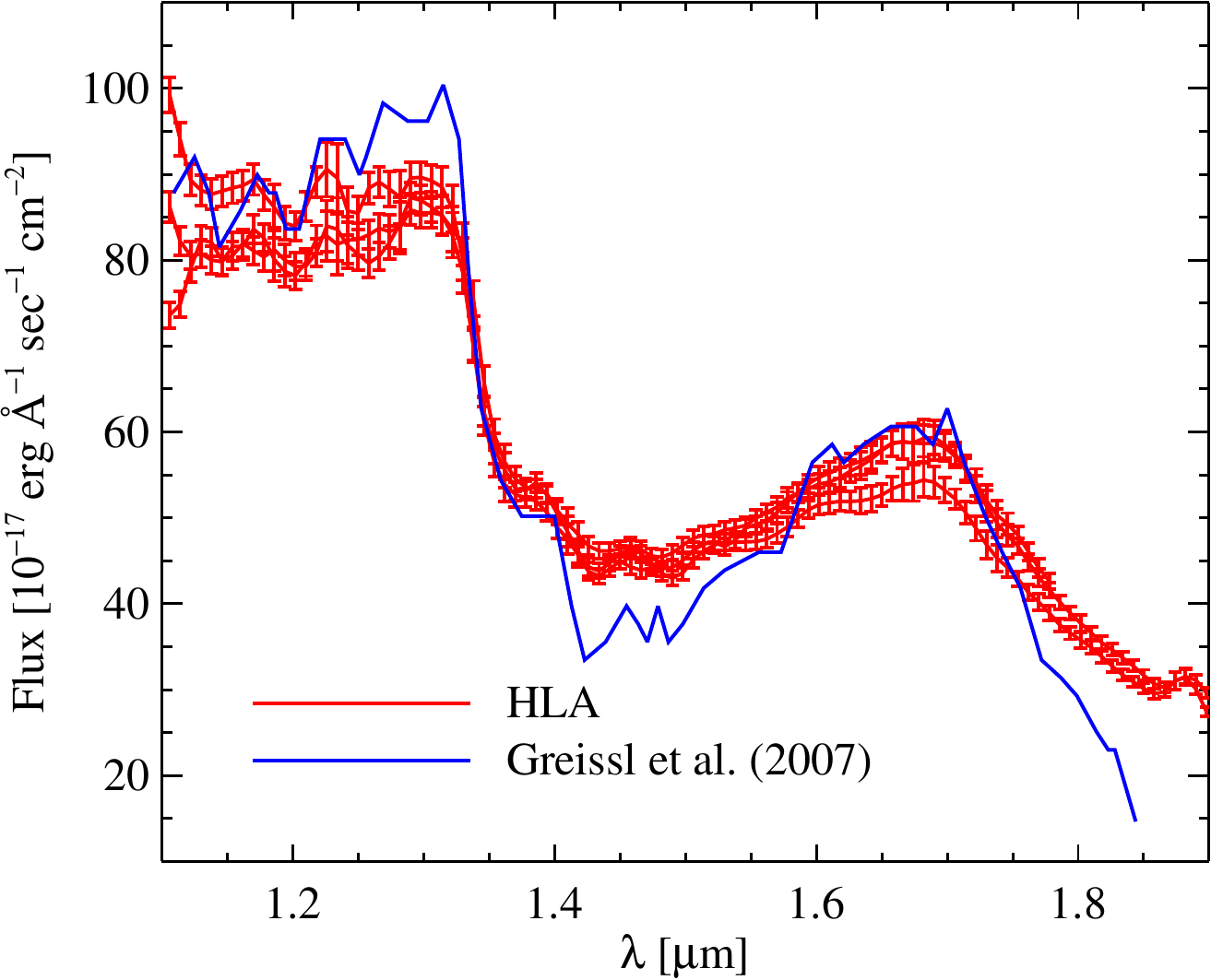}
\caption{Comparison of an HLA spectra of brown dwarf ASR~24 (red curves and
error bars) with the spectrum published by \cite{asr24} (blue curve).  The
latter has been scaled to match the mean flux level of the HLA
data}\label{fig:asr24}
\end{figure}

Finally, in Fig.~\ref{fig:asr24}, we compare HLA spectra of the brown dwarf
ASR~24 with the spectra extracted  from the same data set by \cite{asr24}.
ASR~24 has been observed three times with NICMOS G141 with three different roll
angles. The HLA therefore contains three separate spectra of this source,
namely HNG\_J032911.32+311717.5\_N8VM06BEQ, HNG\_J032911.32+311717.5\_N8VM09G5Q
and HNG\_J032911.32+311717.6\_N8VM16S3Q. \cite{asr24} did not derive a flux
scale for their spectrum, this spectrum is therefore scaled to matched the HLA
extraction. It can be seen that the absolute and relative flux calibration of
the three HLA extractions agree to within 10\%.  We therefore conclude that the
quality of the HLA NICMOS G141 spectra is comparable to the best extractions
previously obtained from the data.

\section{Summary and Conclusion}

The HLA NICMOS grism project provides a database of low resolution H-band IR
spectra. Most of the spectra have never been previously extracted from the HST
NICMOS grism data. The database is useful for work on cool stars and emission
line galaxies at $z$ between 1.1 and 1.9, which are readily detected with the
NICMOS grism. The calibration of the spectra is based on a new analysis of
available calibration data and extraction of the spectra for isolated point
source should be close to optimum.  Confused spectra and spectra of extended
objects are identified in the data release and are also of high quality.  The
absolute and relative flux calibration of the spectra is better than 10\%, and
the wavelength calibration better than 5 nm.  The data release is accompanied
by a wide range of auxiliary data and is available through several interfaces.
We anticipate a revised data release based on a new version of the STScI
CALNICA pipeline in early 2009.

\begin{acknowledgements}

This paper is based on observations made with the NASA/ESA Hubble Space
Telescope, obtained from the data archive at the Space Telescope -- European
Coordinating Facility.  We thank our HLA collaborators Brad Whitmore and  the
STScI and CADC HLA teams.

\end{acknowledgements}

\clearpage

\appendix
\newcommand{\diff}{\mathrm{d}}

\section{Extraction of Slitless Spectra}
\label{sec:slitl-meas}

A crucial difference between longslit and slitless spectroscopy is the
selection of the light to be dispersed. For slitless spetroscopy, the light
that enters the spectrograph is only limited by the  object shape and the point
spread function.  For the extraction, we neglected the fact that the shape, as
well as the point spread function, can change with wavelength. For the
following discussion, we also neglect the impact of the sensitivity curves on
the spectrum. A more rigorous discussion of the parameters will be presented by
\citet{theorypaper}. With the above  approximations, the specific intensity of
the object $I(\vec x, \lambda)$ at the angular position $\vec x$ can be
expressed as the product of two functions

\begin{equation} \label{eq:1} I(\vec x, \lambda) = I(\vec x) F(\lambda) \; ,
\end{equation}
where $I(\vec x)$, is the integrated intensity of the object as revealed by the
direct image, and $F(\lambda)$ is the spectral energy distribution.
$F(\lambda)$ is independent of the position $\vec x$.
The image $\tilde I(\vec x)$ generated by the slitless spectrograph
is then
\begin{equation}
  \label{eq:2}
  \tilde I(\vec x) = \int I \bigl( \vec x - \vec r (\lambda -
  \lambda_0), \lambda \bigr) \, \diff \lambda = \int I \bigl( \vec x -
  \vec r (\lambda - \lambda_0) \bigr) F(\lambda) \, \diff \lambda \; ,
\end{equation}
where $\vec r$ is the dispersion vector and $\lambda_0$ a pivot
wavelength.  The second equality shows that the result can be written
as a simple convolution, carried out along the dispersion direction,
of the original image with the object spectrum.

The goal of the spectral extraction is to add all flux values of pixels that
represent the same wavelength. For slitless spectroscopy, the monochromatic
light from a single wavelength covers a region  with the same shape as the
target object on the detector. For simplicity, we assign all pixels on a
straight line on the two-dimensional grism image to the same wavelength bin in
the one-dimensional spectrum.  The size of the objects along the dispersion
direction limits the spectra resolution of slitless spectra and has the same
effect as the width of the slit in a longslit spectrograph.  In the following,
we will refer to the box defined by the above straight line of constant
wavelength and the size of the object along the dispersion direction  as the
``virtual slit''.

For complex objects, it is unavoidable that a straight line includes pixels
with flux originating from different wavelengths. The goal of the extraction
procedure is to choose the direction of the virtual slit so that variations of
wavelength are minimised.  Our algorithm is designed to work exactly for the
case of elliptical objects, i.e.  in that case each wavelength bin of the
one-dimensional spectrum includes only pixels with flux from a single
wavelength.

\begin{figure}
  \centering
  \includegraphics[scale=0.75]{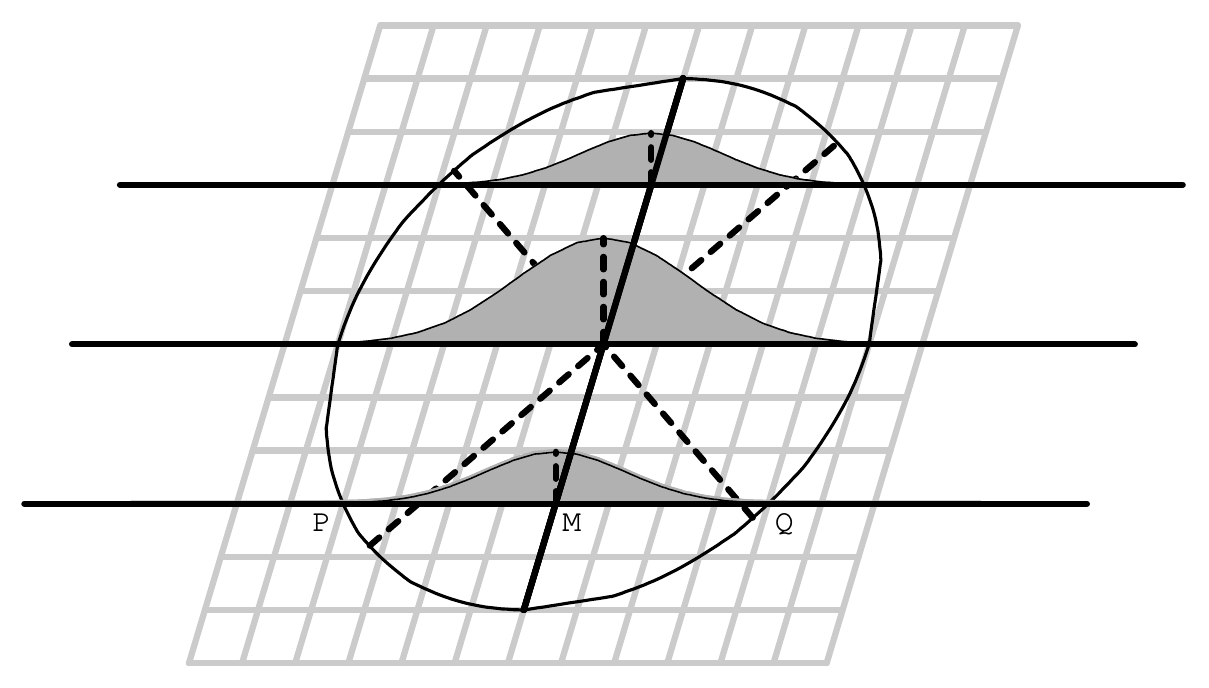}
  \caption{Schematic drawing of a slitless spectrum measurement.}
  \label{fig:angles}
\end{figure}

As shown above in Eq.~\eqref{eq:2}, the two-dimensional spectrum is a
convolution of the object image with the one-dimensional spectrum.  Spectral
features will be smoothed by the object intensity profile corresponding to each
line of dispersion.  For example, for an object with a single emission line at
$\lambda_\mathrm{e}$ so that $F(\lambda) \propto \delta(\lambda -
\lambda_\mathrm{e})$,  the resultant image $\tilde I(\vec x)$ will be identical
to $I(\vec x)$ except for a shift along $\vec r$. 

A natural choice for the orientation of the virtual slit is the direction of
the line defined by the  ``centre of light'' of the intensity profile along the
dispersion direction.  In this case,   the position $\vec x_\mathrm{e}$
satisfies the equation

\begin{equation}
  \label{eq:3}
  \int \tilde I(\vec x_\mathrm{e} + \vec r \ell) \ell \, \diff \ell =
  0 \; .
\end{equation}

For elliptical isophotes, the image of the object can be written as a function
of the elliptical radius $\rho^2 \equiv \vec x^\mathrm{T} A \vec x$, 
\begin{equation}
  \label{eq:4}
  I(\vec x) = f \bigl( \vec x^\mathrm{T} A \vec x\bigr) \; ,
\end{equation}
where $A$ is a $2 \times 2$ symmetric matrix, and where, for simplicity,
we have chosen the origin of our coordinates at the centre of the
elliptical object.  In the following, without loss of generality, we
will also assume that the dispersion direction is along the horizontal
axis $x_1$, so that $\vec r = (r, 0)$.
For convenience, we convert to  a \textit{slanted\/}
coordinate system $\vec x'$ defined by the linear
transformation
\begin{align}
  \label{eq:5}
  x_1 \mapsto x'_1 = {} & x_1 + k x_2 \; , &
  x_2 \mapsto x'_2 = {} & x_2 \; . \\
  \vec x \mapsto \vec x' = {} & T \vec x \; , & 
  T = {} & \begin{pmatrix}
    1 & k \\
    0 & 1
  \end{pmatrix} \; .
\end{align}
In this new coordinate system we then define the slanted direct image
$I'(\vec x')$ as a simple remapping of $I(\vec x)$:
\begin{equation}
  \label{eq:6}
  I'(\vec x') = I(\vec x) = I(T^{-1} \vec x') \; .
\end{equation}
Along a line parallel to the dispersion direction, the transformation
$T$ is a simple translation: in other words, for fixed $x_2$, $x'_1$
is a simple shift of $x_1$ by $k x_2$. 

To compute the orientation of the virtual slit, we choose the 
value of $k$ in the  transformation $T$
in such a manner that the image of the galaxy in the slanted coordinate
system appears as an ellipse with one of the axes oriented along the
dispersion direction.  This requirement is equivalent to choosing the
matrix $T^{-\mathrm{T}} A T^{-1}$ to be  diagonal, which is guaranteed if
$k = A_{12} / A_{11}$.  With this choice, we find
\begin{equation}
  \label{eq:7}
  T^{-\mathrm{T}} A T^{-1} =
  \begin{pmatrix}
    A_{11} & 0 \\
    0 & \det (A) / A_{11}
  \end{pmatrix} \; .
\end{equation}

In this new coordinate system  the virtual slits  are perpendicularly to the
dispersion direction, because the object is now symmetric along the axis
$x'_2$.  The corresponding lines in the original coordinate system can be
obtained by transforming back the vertical lines, and as graphically
illustrated in Fig.~\ref{fig:angles}, these lines can be obtained by joining
the two tangent points of the ellipse describing the direct object with lines
parallel to the dispersion direction.  

Note that the direction of the virtual slits is never along one of the axes of
the elliptical isophotes except in the trivial case where the object is already
oriented along the dispersion direction ($A_{12} = k = 0$).  The angle with
respect to the vertical formed by the lines of constant wavelength is given by
$\tan \beta = k = A_{12} / A_{11}$, and thus for highly elongated objects it
can approach $\pm \pi/2$.

\def\aap{A\& A}
\def\pasp{Proc.Astr.Soc.Pacific}
\def\mnras{MNRAS}
\def\apj{ApJ}
\def\aj{AJ}
\def\apjl{ApJL}
\def\apjs{ApJS}
\def\procspie{Proc. SPIE}

\end{document}